\documentclass[aps,pra,10pt,twocolumn,longbibliography,nofootinbib,floatfix,superscriptaddress]{revtex4-2}
\usepackage{graphicx}
\usepackage{amsmath,amssymb,amsthm,mathtools,bm}
\usepackage{booktabs}
\usepackage{array}
\usepackage{tabularx}
\usepackage{multirow}
\usepackage{xcolor}
\usepackage{microtype}
\usepackage{enumitem}
\usepackage{placeins}
\usepackage{physics}
\usepackage{dsfont}
\usepackage[colorlinks=true,citecolor=blue,linkcolor=blue,urlcolor=blue]{hyperref}
\usepackage[capitalize]{cleveref}
\graphicspath{{fig/}}
\newtheoremstyle{apsthm}%
{\topsep}
{\topsep}
{\normalfont}
{\parindent}
{\itshape}
{.}
{ }
{\thmname{#1}\thmnumber{ #2}\thmnote{ (#3)}}

\theoremstyle{apsthm}
\newtheorem{theorem}{Theorem}


\usepackage{newfloat}
\DeclareFloatingEnvironment[name=Algorithm,placement=t]{algorithm}
\usepackage{algpseudocode}
\algrenewcommand\algorithmicrequire{\textbf{Input:}}
\algrenewcommand\algorithmicensure{\textbf{Output:}}
\algrenewcommand\alglinenumber[1]{\footnotesize #1}
\newcommand{\One}{\mathbf{1}}

\begin{document}

\title{Binary-Tree Quantum Approximate Optimization Algorithm for the Maximum $k$-Cut Problem}

\author{Zhaoyang Song}
\affiliation{Center for Quantum Sciences and School of Physics, Northeast Normal University, Changchun 130024, China}
\author{Wei Zhang}
\affiliation{Department of Physics, College of Science, Yanbian University, Yanji, Jilin 133002, China}
\affiliation{Institute of Quantum Science and Technology, Yanbian University, Yanji, Jilin 133002, China}
\author{Ji Cao}
\email{caoji@ybu.edu.cn}
\affiliation{Department of Physics, College of Science, Yanbian University, Yanji, Jilin 133002, China}
\affiliation{Institute of Quantum Science and Technology, Yanbian University, Yanji, Jilin 133002, China}
\author{Xuexi Yi}
\email{yixx@nenu.edu.cn}
\affiliation{Center for Quantum Sciences and School of Physics, Northeast Normal University, Changchun 130024, China}
\author{Hong-Fu Wang}
\email{hfwang@ybu.edu.cn}
\affiliation{Center for Quantum Sciences and School of Physics, Northeast Normal University, Changchun 130024, China}
\affiliation{Department of Physics, College of Science, Yanbian University, Yanji, Jilin 133002, China}
\affiliation{Institute of Quantum Science and Technology, Yanbian University, Yanji, Jilin 133002, China}
\affiliation{Institute for Interdisciplinary Quantum Information Technology, Jilin Engineering Normal University, Changchun, Jilin 130052, China}
\affiliation{Jilin Engineering Laboratory for Quantum Information Technology, Changchun, Jilin 130052, China}

\date{\today}

\begin{abstract}
The maximum $k$-cut (Max-$k$-Cut) problem, as a natural generalization of the maximum-cut (MaxCut) problem, has broader applications in practical engineering. To solve this problem on quantum devices, existing studies typically encode the $k$ possible subset assignments of each vertex within the quantum approximate optimization algorithm (QAOA), which directly causes the required quantum resources to grow substantially with $k$. Here we propose a binary-tree quantum approximate optimization algorithm (BT-QAOA), which solves the Max-$k$-Cut problem by applying standard MaxCut QAOA to at most $2k-3$ subproblems. Specifically, after solving MaxCut once on the full vertex set, each subsequent step splits only the subset that yields the largest cut-value gain. This process is equivalent to constructing a binary tree whose leaf nodes after $k-1$ steps form an approximate solution to the Max-$k$-Cut problem. Furthermore, we derive a theoretical lower bound on the approximation ratio in terms of the single-step splitting quality, showing that the overall approximation ratio does not decay exponentially across multiple splits. Numerical simulations show that BT-QAOA substantially reduces the single-circuit qubit and gate overhead while maintaining a high approximation ratio, improving the estimated success probability of a single compiled circuit by orders of magnitude. This work effectively expands the practical reach of QAOA on near-term quantum hardware, offering a new route for solving more complex real-world optimization problems with limited quantum resources.
\end{abstract}

\maketitle

%
\section{Introduction}\label{sec:intro}

The maximum $k$-cut (Max-$k$-Cut) problem is an NP-hard problem in combinatorial optimization. Its better-known special case at $k=2$ is the maximum-cut (MaxCut) problem, which has been proven to be NP-complete~\cite{Karp1972}. This problem corresponds to graph partitioning in graph theory, namely partitioning the vertices of a given graph into $k$ disjoint subsets so as to maximize the total weight of edges spanning different subsets~\cite{Frieze1997,deKlerk2004}. Practical problems of broad significance, such as frequency allocation in wireless communications~\cite{Eisenblaetter2002}, scheduling~\cite{Carlson1966}, modular product design~\cite{Gaur2008,Fuchs2021,Fuchs2025}, and network design~\cite{Apte2026a}, are all typical applications.

The quantum approximate optimization algorithm (QAOA) is a representative algorithm for solving MaxCut on near-term quantum computers~\cite{Farhi2014,Zhou2020,Harrigan2021,He2025}. In certain regimes, it has also demonstrated theoretical advantages over specific classical approximation algorithms~\cite{Apte2026a}. As research on MaxCut continues to mature~\cite{Wang2018,Wurtz2021}, a natural extension is to apply QAOA to explore the more general Max-$k$-Cut problem.

In quantum computing, the main difference between Max-$k$-Cut and MaxCut lies in the encoding scheme. A single qubit has only two energy levels, matching the two subset assignments in MaxCut. Therefore, for the $k$ possible assignments in Max-$k$-Cut, a new encoding scheme must be designed. One-hot encoding uses $k$ qubits to represent a single vertex, keeping interactions in two-qubit form, yet the total number of qubits grows linearly with $k$~\cite{Lucas2014,Hadfield2019,WangXY2020,Harkness2025}. By contrast, binary encoding reduces the number of qubits required per vertex to $\lceil\log_2 k\rceil$. However, representing a single vertex with multiple qubits introduces multiqubit interactions for each edge, increasing circuit depth and accumulated error after decomposition~\cite{Fuchs2021,Fuchs2025}. High-dimensional quantum schemes use $k$-level quantum units (qudits) so that a single vertex corresponds to $k$ subsets~\cite{Deller2023,Bravyi2022,Apte2026a}. Yet they rely on high-dimensional quantum hardware, where coherent control of multilevel states and high-fidelity gate implementations still face substantial technical challenges~\cite{Ringbauer2022}. Although these schemes differ in their specific forms, they all focus on encoding all $k$ possibilities in the quantum state of each vertex. While this idea is intuitive, it inevitably incurs substantial resource or hardware overhead. This raises the question of whether one can move beyond the framework of multistate encoding and solve the Max-$k$-Cut problem from the perspective of algorithm design.

To address this question, we propose a binary-tree quantum approximate optimization algorithm (BT-QAOA) for the Max-$k$-Cut problem. Rather than directly encoding $k$ subsets in a single quantum circuit, this method starts from the full vertex set of the original graph and obtains $k$ subsets step by step through $k-1$ bipartitions. Specifically, the algorithm first uses QAOA to solve MaxCut on the original graph, partitioning the full vertex set into two subsets. Each subsequent step solves MaxCut on the induced subgraph of each subset and selects the subset with the largest cut-value gain for further bipartitioning. The entire partitioning process is equivalent to constructing a binary tree rooted at the original vertex set with the final $k$ subsets as its leaf nodes. Throughout this sequential partitioning process, regardless of how $k$ increases, the quantum device always solves the standard MaxCut problem. This avoids the sharp growth in single-circuit resource overhead with $k$ seen in direct encoding schemes. Meanwhile, by recursively analyzing the cut-value gain at each step, we establish a quantitative relation between the single-step splitting quality and the final Max-$k$-Cut approximation ratio, with the resulting lower bound gradually approaching $1$ as $k$ increases. We also perform numerical simulations of both solution accuracy and hardware compilation overhead. Compared with direct encoding schemes, BT-QAOA substantially reduces the number of qubits and two-qubit gates per circuit while maintaining a high approximation ratio. Through a structured decomposition of the Max-$k$-Cut problem, our work transforms a problem that originally required complex encodings into a sequence of standard MaxCut subproblems requiring only a single qubit per vertex. This lowers the hardware resource threshold while preserving solution quality, providing a viable path for near-term quantum devices to solve multiway graph partitioning problems.

%
\section{Preliminaries}\label{sec:preliminaries}

\subsection{The Max-$k$-Cut problem}\label{subsec:max_k_cut}

\begin{figure}[!htbp]
\centering
\includegraphics[width=0.82\columnwidth]{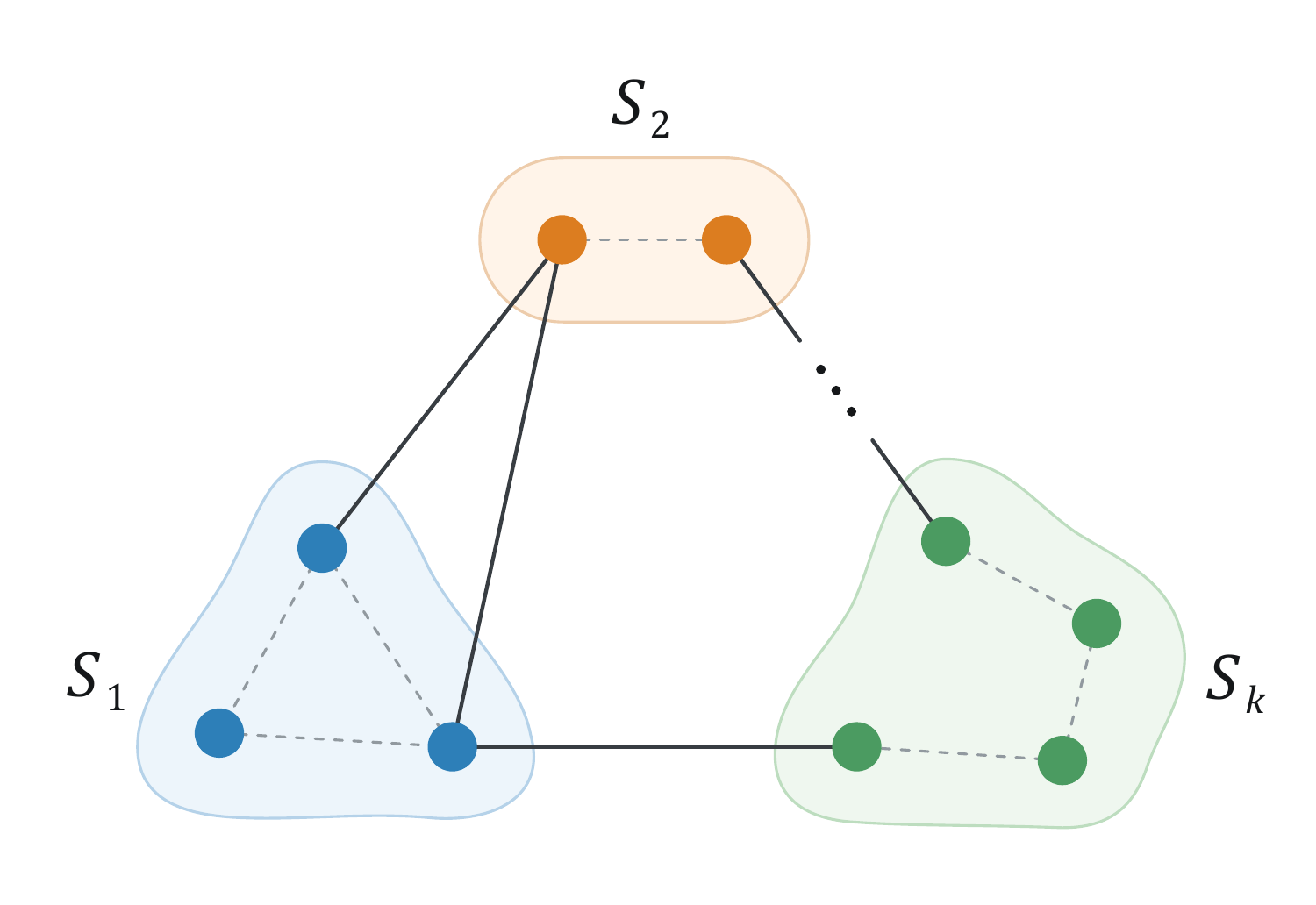}
\caption{Illustration of the Max-$k$-Cut problem. The vertices of a graph are partitioned into $k$ mutually disjoint subsets $S_1,S_2,\dots,S_k$. Only cut edges connecting different subsets (solid lines) contribute to the objective function $C(c)$, whereas edges within a subset (dashed lines) contribute $0$.}\label{fig:encoding}
\end{figure}

Let $G=(V,E,w)$ be a finite undirected weighted graph, where $|V|=n$, $|E|=m$, and each undirected edge $\{i,j\}\in E$ has a nonnegative weight $w_{ij}\ge 0$. As shown in Fig.~\ref{fig:encoding}, given a number of groups $2\le k\le n$, the Max-$k$-Cut problem requires partitioning the vertex set $V$ into $k$ mutually disjoint subsets $S_1,S_2,\dots,S_k$ so as to maximize the total weight of edges connecting different subsets. Let the label vector $c=(c_1,\dots,c_n)\in\{1,\dots,k\}^n$ represent the group assignment of each vertex. The cut value of partition $c$ is defined as
\begin{equation}\label{eq:max_k_cut_obj}
C(c)=\sum_{\{i,j\}\in E} w_{ij}\,\One[c_i\ne c_j],
\end{equation}
where the indicator function $\One[c_i\ne c_j]$ takes the value $1$ if and only if $c_i\ne c_j$, and $0$ otherwise. On an unweighted graph with $w_{ij}\equiv 1$, $C(c)$ equals the number of cut edges. Denote the global maximum cut value attainable over all valid $k$-partitions by
\begin{equation}\label{eq:opt_k_def}
C_{\max}=\max_{c\in\{1,\dots,k\}^n} C(c).
\end{equation}
Because different graph instances vary substantially in size and edge weights, the raw cut value $C(c)$ is not suitable for direct comparison across instances. We therefore define the approximation ratio $\eta=C(c)/C_{\max}$ to measure the solution quality of partition $c$ relative to the global optimum.

\subsection{QAOA for MaxCut}\label{subsec:qaoa_maxcut}

When $k=2$, Max-$k$-Cut reduces to the standard MaxCut problem. In this case, each vertex only needs to distinguish between two subset assignments, so the $n$ vertices can be directly represented by $n$ logical qubits. The outcomes $0$ and $1$ of computational-basis measurements in $\{|0\rangle,|1\rangle\}$ correspond to the two subsets. Let $Z_i$ denote the Pauli-$Z$ operator acting on the $i$th qubit. For an edge $\{i,j\}\in E$, the operator $\frac{1}{2}(I-Z_i Z_j)$ has eigenvalue $1$ when its two endpoints are in different computational-basis states and eigenvalue $0$ when they are in the same computational-basis state. Accordingly, the bipartitioning objective in Eq.~\eqref{eq:max_k_cut_obj} maps directly to the MaxCut cost Hamiltonian
\begin{equation}\label{eq:maxcut_hamiltonian}
\hat{C}=\sum_{\{i,j\}\in E}\frac{w_{ij}}{2}(I-Z_i Z_j).
\end{equation}

QAOA increases the expectation value of the cost Hamiltonian $\hat{C}$ by optimizing a parameterized quantum circuit, thereby obtaining measurement samples with high cut values~\cite{Farhi2014,Blekos2024}. The algorithm introduces the transverse-field mixing Hamiltonian $\hat{B}=\sum_{i\in V} X_i$, takes the uniform superposition $|+\rangle^{\otimes n}$ as the initial state, and alternates $p$ layers of cost and mixing evolutions parameterized by $\boldsymbol{\theta}=(\gamma_1,\dots,\gamma_p,\beta_1,\dots,\beta_p)\in\mathbb{R}^{2p}$ to prepare
\begin{equation}\label{eq:qaoa_state}
|\psi(\boldsymbol{\theta})\rangle = e^{-\mathrm{i}\beta_p \hat{B}}e^{-\mathrm{i}\gamma_p \hat{C}}\cdots e^{-\mathrm{i}\beta_1 \hat{B}}e^{-\mathrm{i}\gamma_1 \hat{C}}|+\rangle^{\otimes n},
\end{equation}
where $X_i$ is the Pauli-$X$ operator on the $i$th qubit. In the circuit implementation, up to a global phase, the cost evolution $e^{-\mathrm{i}\gamma_h \hat{C}}$ in layer $h=1,\dots,p$ decomposes into two-qubit rotation gates $R_{ZZ}(-\gamma_h w_{ij})=\exp(\mathrm{i}\frac{\gamma_h w_{ij}}{2}Z_i Z_j)$ on each edge. The mixing evolution $e^{-\mathrm{i}\beta_h \hat{B}}$ decomposes into single-qubit rotation gates $R_X(2\beta_h)=\exp(-\mathrm{i}\beta_h X_i)$ on each qubit. During variational training, the algorithm repeatedly performs computational-basis measurements at the current parameters $\boldsymbol{\theta}$ and estimates the expectation value $\langle\psi(\boldsymbol{\theta})|\hat{C}|\psi(\boldsymbol{\theta})\rangle$ from the sample mean cut value. A classical optimizer then iteratively updates $\boldsymbol{\theta}$ to maximize this estimate. After training converges, the optimized parameters $\boldsymbol{\theta}^*$ are fixed to sample a new batch of bitstrings, from which the bitstring with the highest cut value is selected as the output partition.

\subsection{Quantum encodings of Max-$k$-Cut}\label{subsec:quantum_encodings}

One-hot encoding assigns $k$ logical qubits to each vertex $i\in V$. Binary variables $x_{ia}\in\{0,1\}$ denote the corresponding computational-basis values, where $x_{ia}=1$ indicates that vertex $i$ belongs to the $a$th subset. The full graph requires $kn$ logical qubits in total, and each vertex must satisfy the constraint $\sum_{a=1}^{k} x_{ia}=1$. Within the subspace satisfying this constraint, the indicator function for the two endpoints of edge $\{i,j\}$ belonging to the same group is $\sum_{a=1}^{k} x_{ia} x_{ja}$, so each edge corresponds to $k$ two-body interaction terms in each cost layer. To exclude invalid encoded states that violate the constraint, one approach is to prepare the initial state within the feasible subspace and evolve it with a constraint-preserving XY mixer~\cite{Hadfield2019,WangXY2020}. Another approach, under a transverse-field mixer, is to subtract from the maximization objective a quadratic penalty term $\lambda\sum_{i\in V}(\sum_{a=1}^{k} x_{ia}-1)^2$ with penalty coefficient $\lambda>0$~\cite{Lucas2014}.

Binary encoding assigns $L=\lceil\log_2 k\rceil$ logical qubits to each vertex $i\in V$~\cite{Fuchs2021}, reducing the total logical qubit count of the graph to $nL$. Labeling the $k$ subsets as $0,\dots,k-1$, each vertex is encoded by the $L$-bit binary representation of its group index. Let $Z_{i\ell}$ denote the Pauli-$Z$ operator on the $\ell$th qubit of vertex $i$, where $\ell=1,\dots,L$. When $k=2^L$, two vertices belong to the same group if and only if all $L$ corresponding bits take the same values, so the same-group indicator operator is
\begin{equation}\label{eq:binary_same_group}
\One[c_i=c_j]=\prod_{\ell=1}^{L}\frac{I+Z_{i\ell}Z_{j\ell}}{2},
\end{equation}
and the corresponding single-edge cost term is $w_{ij}(I-\One[c_i=c_j])$. Expanding Eq.~\eqref{eq:binary_same_group} yields $2^L-1$ nonidentity Pauli-$Z$ product terms, with the highest-order term acting on $2L$ qubits. These many-body terms must be further decomposed into elementary gates supported by the hardware. When $k$ is not a power of $2$, the extra $2^L-k$ encoded states must also be handled, for instance by introducing penalty terms, restricting the evolution subspace, or merging the extra encoded states into existing groups. In addition, high-dimensional quantum representations encode each vertex using a single $k$-level quantum unit~\cite{Deller2023}, but rely on corresponding high-dimensional quantum hardware.

%
\section{The BT-QAOA method}\label{sec:method}

\subsection{Binary-tree representation and problem transformation}\label{subsec:binary_tree}

\begin{figure*}[t]
\centering
\includegraphics[width=0.96\textwidth]{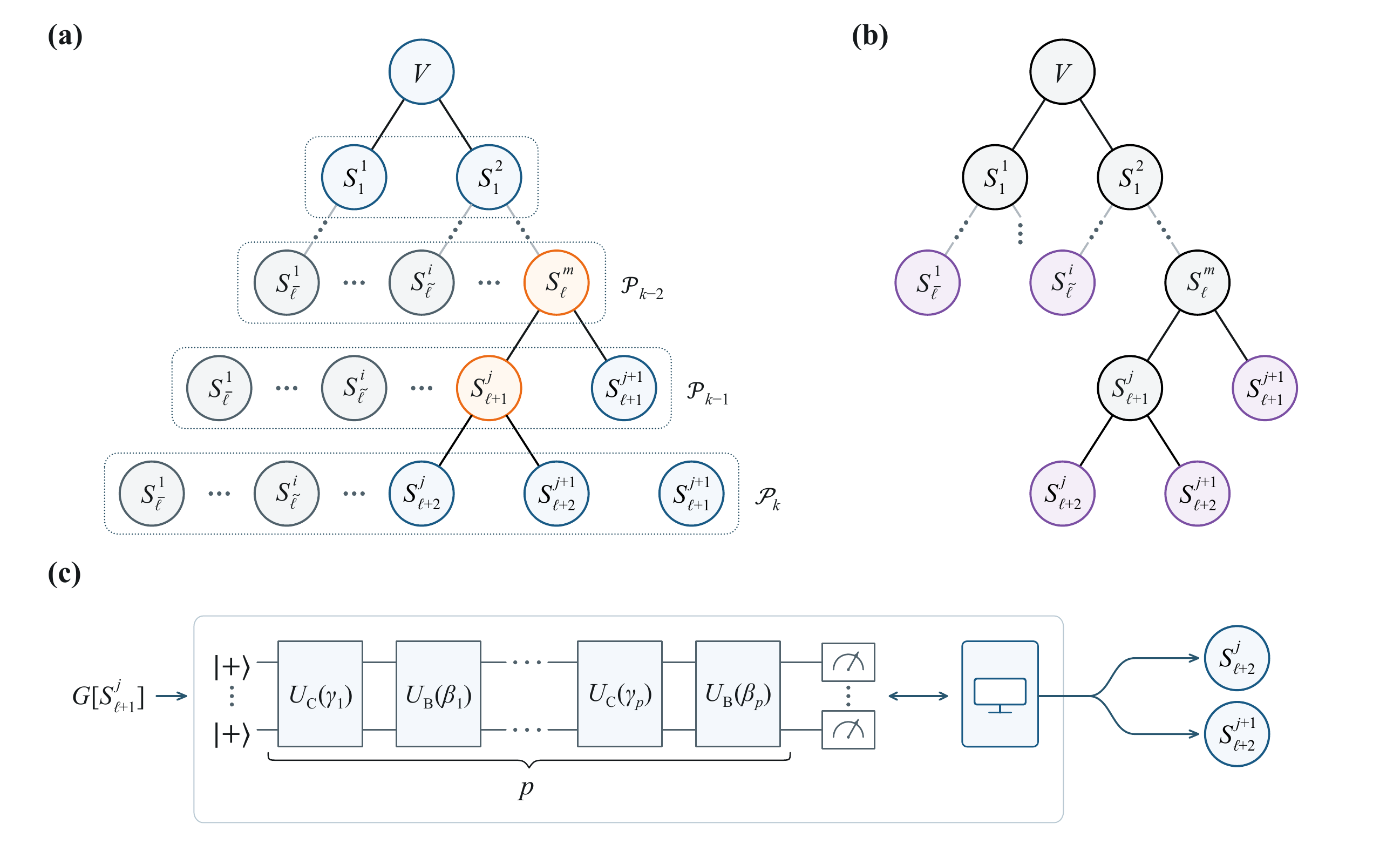}
\caption{Schematic of the splitting process, binary-tree structure, and single-node QAOA solver in BT-QAOA. (a) Forming $k$ subsets from the full vertex set $V$ through $k-1$ splits. Blue dashed boxes indicate the current partition $\mathcal{P}_t$, orange nodes indicate the subset selected for further splitting at each step, and gray nodes indicate retained subsets, whose tree depths $\bar{\ell}$, $\tilde{\ell}$, and $\ell$ need not be equal. (b) Corresponding binary-tree representation, where black nodes are internal nodes that have been split and purple leaf nodes at different depths form the final partition $\mathcal{P}_k$. (c) The $p$-layer MaxCut QAOA solving process for a leaf node on its corresponding induced subgraph.}\label{fig:btqaoa_flow}
\end{figure*}

Consider the case of Max-$4$-Cut. First, solving MaxCut once on the original graph yields two subsets; then, successively solving MaxCut on subsets that can be further split produces an approximate solution to Max-$4$-Cut after three splits. Following this idea, for larger $k$, an approximate solution to Max-$k$-Cut can likewise be constructed step by step through multiple splits as shown in Fig.~\ref{fig:btqaoa_flow}(a). This process is equivalent to constructing the binary tree shown in Fig.~\ref{fig:btqaoa_flow}(b).

Each node in the tree represents a vertex subset $S_d^r$ of the original graph, where the subscript $d$ denotes the tree depth and the superscript $r$ indexes the nodes at the same depth. The algorithm starts from the initial partition $\mathcal{P}_1=\{V\}$ corresponding to the root node. Suppose that before the $t$th split, the current partition $\mathcal{P}_t$ contains $t$ subsets. The $t$th split selects one subset $S$ from $\mathcal{P}_t$ and divides it into two mutually disjoint nonempty subsets $S_0$ and $S_1$, while the remaining $t-1$ subsets are kept unchanged. This yields the new partition $\mathcal{P}_{t+1}=(\mathcal{P}_t\setminus\{S\})\cup\{S_0,S_1\}$. After $k-1$ splits, all $k$ leaf nodes distributed across different depths of the tree together form the final partition $\mathcal{P}_k$.

Splitting a leaf node $S\in\mathcal{P}_t$ into $(S_0,S_1)$ does not change the status of existing cut edges across subsets. All newly cut edges come from the induced subgraph $G[S]=(S,E(S))$ of $S$, where $E(S)=\{\{u,v\}\in E \mid u,v\in S\}$. Therefore, the single-step increase in the cut value of the full graph equals the cut value of the split $(S_0,S_1)$ on the subgraph $G[S]$,
\begin{equation}\label{eq:bisection_gain}
g(S)=\sum_{\substack{u\in S_0,\,v\in S_1\\\{u,v\}\in E(S)}} w_{uv}.
\end{equation}
Maximizing this single-step increment is equivalent to solving the standard MaxCut problem on the subgraph $G[S]$.

\subsection{QAOA-based node splitting}\label{subsec:node_qaoa}

As shown in Fig.~\ref{fig:btqaoa_flow}(c), for a leaf node $S$ that requires further splitting, the algorithm runs the $p$-layer standard MaxCut QAOA of Sec.~\ref{subsec:qaoa_maxcut} on its induced subgraph $G[S]=(S,E(S))$. The number of qubits used equals the number of vertices $|S|$ in the subset. Once variational parameter training is completed, the bitstring with the highest cut value is selected from the final measurement samples. According to the measurement outcomes $0$ and $1$ of each vertex, $S$ is split into two subsets $S_0$ and $S_1$, and the corresponding cut-value gain $g(S)$ is calculated via Eq.~\eqref{eq:bisection_gain}.

\subsection{Node selection and algorithm procedure}\label{subsec:greedy_selection}

Before the $t$th split for $t\ge 2$, the current partition $\mathcal{P}_t$ contains $t$ leaf nodes. BT-QAOA compares the cut-value gains $g(S)$ among the leaf nodes satisfying $|S|\ge 2$ and selects the leaf node $S^*$ with the largest cut-value gain for splitting according to
\begin{equation}\label{eq:greedy_choice}
S^*=\arg\max_{\substack{S\in\mathcal{P}_t\\|S|\ge 2}} g(S),
\end{equation}
which corresponds to the orange node in Fig.~\ref{fig:btqaoa_flow}(a). Replacing $S^*$ in $\mathcal{P}_t$ with $S^*_0$ and $S^*_1$ then yields $\mathcal{P}_{t+1}$, and the cumulative cut value is updated as $C_{t+1}=C_t+g(S^*)$.

\begin{algorithm}[!htbp]
\caption{Binary-tree quantum approximate optimization algorithm for Max-$k$-Cut}\label{alg:btqaoa}
\begin{algorithmic}[1]
\Require Nonnegative-weight undirected graph $G=(V,E,w)$, target number of groups $2\le k\le n$, node QAOA depth $p$, and training and sampling budgets
\Ensure Partition $\mathcal{P}_k$ into $k$ subsets and total cut value $C_k$
\State Initialize current partition $\mathcal{P}_1 \gets \{V\}$, current subset count $t \gets 1$, and cumulative cut value $C_1 \gets 0$
\State Initialize record table $\mathcal{T} \gets \varnothing$
\While{$t < k$}
    \For{each subset $S\in\mathcal{P}_t$ with $|S|\ge 2$ not yet stored in $\mathcal{T}$}
        \State Run $p$-layer MaxCut QAOA on subgraph $G[S]$ to obtain $(S_0,S_1)$ and cut-value gain $g(S)$
        \State Store $(S_0,S_1,g(S))$ in $\mathcal{T}$
    \EndFor
    \State Select $S^*$ from $\mathcal{P}_t$ according to Eq.~\eqref{eq:greedy_choice}
    \State Retrieve $(S^*_0, S^*_1, g(S^*))$ from $\mathcal{T}$ and update $\mathcal{P}_{t+1} \gets (\mathcal{P}_t \setminus \{S^*\}) \cup \{S^*_0, S^*_1\}$
    \State Update cumulative cut value $C_{t+1} \gets C_t + g(S^*)$ and set $t \gets t + 1$
\EndWhile
\State \Return Final partition $\mathcal{P}_k$ and total cut value $C_k$
\end{algorithmic}
\end{algorithm}

Because unselected leaf nodes and their induced subgraphs $G[S]$ remain unchanged upon entering the next step, the algorithm reuses their previously obtained splitting results through the record table $\mathcal{T}$. If the total number of subsets has not yet reached $k$, the next step only needs to solve MaxCut on the two newly generated child nodes. The complete procedure is given in Algorithm~\ref{alg:btqaoa}.

%
\section{Quantum resource analysis}\label{sec:resources}

\subsection{Resource overhead of BT-QAOA}\label{subsec:btqaoa_resources}

Let the induced subgraph $G[S]=(S,E(S))$ of a leaf node $S$ contain $n_S=|S|$ vertices and $m_S=|E(S)|$ edges. For a single run of $p$-layer standard MaxCut QAOA, the circuit requires $n_S$ logical qubits. In a single cost-evolution layer, each edge corresponds to a two-qubit rotation gate $R_{ZZ}(-\gamma_h w_{uv})$, whose standard decomposition consists of $2$ logical CNOT gates and $1$ single-qubit $R_Z$ gate. Therefore, a single cost-evolution layer at node $S$ requires $2m_S$ logical CNOT gates in total. The root node corresponds to the original graph $G$ and requires $n$ logical qubits and $2m$ logical CNOT gates per layer. Subsequent child nodes have fewer vertices and no more edges, so the root node sets the single-circuit resource upper bound across all nodes. As seen from Algorithm~\ref{alg:btqaoa}, as the target number of groups $k$ increases, BT-QAOA does not increase the qubit or gate count of a single quantum circuit. Instead, it turns the growth of $k$ into the sequential solution of at most $2k-3$ MaxCut subproblems.

\subsection{Comparison with direct encodings}\label{subsec:encoding_comparison}

\begin{figure}[t]
\centering
\includegraphics[width=\columnwidth]{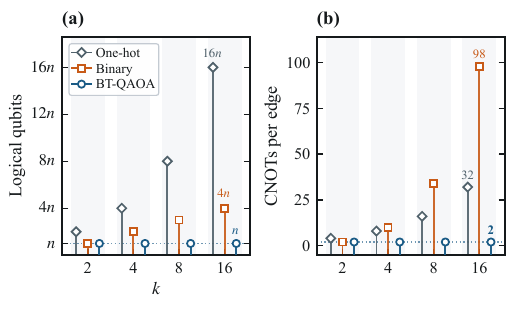}
\caption{Comparison of logical resources among BT-QAOA, one-hot encoding, and binary encoding for solving Max-$k$-Cut on an $n$-vertex graph at discrete subset counts $k\in\{2,4,8,16\}$. (a) Logical qubits required per quantum circuit. (b) Logical CNOT gates required per edge in a single cost-evolution layer under standard independent Pauli-$Z$ decomposition, where the horizontal dotted line marks the $k$-independent constant baseline of BT-QAOA.}\label{fig:resource_comparison}
\end{figure}

We compare BT-QAOA with one-hot encoding~\cite{Hadfield2019,WangXY2020,Lucas2014} and binary encoding~\cite{Fuchs2021} on the same $n$-vertex graph $G=(V,E,w)$, with the logical resources as functions of the target subset count $k$ shown in Fig.~\ref{fig:resource_comparison}. As seen in Fig.~\ref{fig:resource_comparison}(a), for the number of logical qubits required per circuit, one-hot encoding and binary encoding require $nk$ and $n\lceil\log_2 k\rceil$ qubits, respectively, both growing as $k$ increases. By contrast, BT-QAOA uses at most $n$ qubits at the root node and satisfies $n_S<n$ at all subsequent child nodes.

For the number of logical CNOT gates required per edge in a single cost-evolution layer in Fig.~\ref{fig:resource_comparison}(b), BT-QAOA requires a fixed count of $2$, whereas decomposing the cost terms alone in one-hot encoding already requires $2k$ logical CNOT gates. For binary encoding, when $k=2^L$, independently decomposing the $k-1$ Pauli-$Z$ product terms in Eq.~\eqref{eq:binary_same_group} requires $2k(\log_2 k-1)+2$ logical CNOT gates. Even when intermediate parities are reused as in Appendix~\ref{app:resources} to reduce the gate count, a single edge still requires $k+2\log_2 k-2$ logical CNOT gates in a single cost-evolution layer. When $k\ne 2^L$, redundant encoded states must additionally be handled.

%
\section{Approximation guarantee}\label{sec:theory}

\subsection{Approximation guarantee and its derivation}\label{subsec:guarantee_derivation}

At step $t$ of Algorithm~\ref{alg:btqaoa}, the current partition $\mathcal{P}_t$ contains $t$ subsets, and the total weight of the edges already cut is $C_t$, with $C_1=0$ initially. Let $\Delta_t=\max_{S\in\mathcal{P}_t} g(S)$ be the cut-value gain produced by splitting the leaf node $S^*$ selected at step $t$ according to Eq.~\eqref{eq:greedy_choice}. As shown in Sec.~\ref{subsec:binary_tree}, all previously cut edges are retained in subsequent splits, so $C_{t+1}=C_t+\Delta_t$. Because the edge weights are nonnegative, $C_t\le C_{\max}$ always holds for $t\le k\le n$, and the difference between the current cumulative cut value and the global optimum is $C_{\max}-C_t\ge 0$.

To analyze how much the single-step cut-value gain $\Delta_t$ reduces the difference $C_{\max}-C_t$, consider a globally optimal $k$-partition $c^*\in\{1,\dots,k\}^n$ of the original graph with $C(c^*)=C_{\max}$. For each subset $S$ in the current partition $\mathcal{P}_t$, let $R_t(S)$ denote the total weight of edges that are cut in the optimal partition $c^*$ but still remain inside the subgraph $G[S]$ at the current step, namely,
\begin{equation}\label{eq:RtS_def}
R_t(S)=\sum_{\{u,v\}\in E(S)} w_{uv}\,\One[c^*_u\ne c^*_v],
\qquad
R_t=\sum_{S\in\mathcal{P}_t} R_t(S).
\end{equation}
Among all cut edges of the optimal partition $c^*$, apart from the portion $R_t$ remaining inside the subgraphs $G[S]$, the rest have already been cut in the current partition $\mathcal{P}_t$, with total weight $C_{\max}-R_t$. Because these edges form a subset of the currently cut edges and the edge weights are nonnegative, their total weight does not exceed the current cumulative cut value $C_t$, which gives
\begin{equation}\label{eq:Rt_lower_bound}
R_t\ge C_{\max}-C_t.
\end{equation}

We next establish the quantitative relation between $R_t(S)$ and the maximum cut value $M(S)$ of the subgraph $G[S]$~\cite{Samal2009}. Divide the label set $\{1,\dots,k\}$ of the optimal partition randomly and evenly into two classes by choosing $b=\lfloor k/2\rfloor$ labels uniformly at random for the first class and assigning the remaining $k-b=\lceil k/2\rceil$ labels to the second class. This accordingly divides the vertices of subset $S$ into two parts. For any edge $\{u,v\}\in E(S)$ satisfying $c^*_u\ne c^*_v$, the probability that label $c^*_u$ falls into the first class and $c^*_v$ falls into the second class is $\frac{b}{k}\frac{k-b}{k-1}$, and the reverse assignment has the same probability. Therefore, the probability that this edge is cut is
\begin{equation}\label{eq:ak_def}
a_k=\frac{2b(k-b)}{k(k-1)}=\frac{2\lfloor k^2/4\rfloor}{k(k-1)}
=\begin{cases}
\dfrac{k}{2(k-1)}, & k\text{ even},\\[6pt]
\dfrac{k+1}{2k}, & k\text{ odd}.
\end{cases}
\end{equation}
By linearity of expectation, the expected edge weight cut by this random bipartition on $G[S]$ is at least $a_k R_t(S)$, so the maximum cut value $M(S)$ of $G[S]$ must satisfy
\begin{equation}\label{eq:MS_lower_bound}
M(S)\ge a_k R_t(S).
\end{equation}

In actual execution, QAOA obtains an approximate MaxCut on the subgraph $G[S]$. Suppose that at step $t$, the cut-value gain $g(S)$ of each subset in the current partition $\mathcal{P}_t$ relative to the subgraph maximum cut value $M(S)$ satisfies
\begin{equation}\label{eq:uniform_quality_cond}
g(S)\ge r_t M(S),\qquad \forall S\in\mathcal{P}_t,
\end{equation}
where $r_t\in[0,1]$ is the splitting-quality lower bound at step $t$. Because the algorithm selects the subset in $\mathcal{P}_t$ with the largest cut-value gain for splitting, and the maximum is no smaller than the arithmetic mean over all $t$ subsets, combining Eqs.~\eqref{eq:Rt_lower_bound}, \eqref{eq:MS_lower_bound}, and \eqref{eq:uniform_quality_cond} yields
\begin{equation}\label{eq:delta_t_chain}
\begin{aligned}
\Delta_t
&=\max_{S\in\mathcal{P}_t} g(S)
\ge \frac{1}{t}\sum_{S\in\mathcal{P}_t} g(S) \\
&\ge \frac{r_t}{t}\sum_{S\in\mathcal{P}_t} M(S)
\ge \frac{a_k r_t}{t}\sum_{S\in\mathcal{P}_t} R_t(S) \\
&=\frac{a_k r_t}{t}R_t
\ge \frac{a_k r_t}{t}(C_{\max}-C_t).
\end{aligned}
\end{equation}
Substituting $\Delta_t=C_{t+1}-C_t$ into Eq.~\eqref{eq:delta_t_chain} and following the recurrence approach in greedy approximation analysis~\cite{Nemhauser1978}, we obtain the single-step recurrence relation
\begin{equation}\label{eq:gap_recurrence}
C_{\max}-C_{t+1}\le\left(1-\frac{a_k r_t}{t}\right)(C_{\max}-C_t).
\end{equation}
Since $0<a_k\le 1$ and $r_t\in[0,1]$, the coefficient at each step satisfies $0\le 1-a_k r_t/t\le 1$. Iterating the above recurrence relation $k-1$ times from $C_1=0$ yields a lower bound on the final approximation ratio.

\begin{theorem}[Approximation guarantee for BT-QAOA]\label{thm:product_bound}
Let $G=(V,E,w)$ be a nonnegative-weight undirected graph, let the target number of subsets satisfy $2\le k\le |V|$, and suppose that the global Max-$k$-Cut value satisfies $C_{\max}>0$. If at step $t$ of Algorithm~\ref{alg:btqaoa}, all subsets $S\in\mathcal{P}_t$ in the current partition $\mathcal{P}_t$ satisfy $g(S)\ge r_t M(S)$, where $r_t\in[0,1]$ and $M(S)$ is the maximum cut value of the induced subgraph $G[S]$, then the approximation ratio $\eta$ corresponding to the final cut value $C_k$ output by the algorithm satisfies
\begin{equation}\label{eq:product_bound}
\eta\ge 1-\prod_{t=1}^{k-1}\left(1-\frac{a_k r_t}{t}\right),
\end{equation}
where $a_k=\dfrac{2\lfloor k^2/4\rfloor}{k(k-1)}$.
\end{theorem}

\subsection{Effect of splitting quality on the guarantee}\label{subsec:quality_impact_analysis}

If the splitting quality of every leaf node throughout the solving process satisfies a uniform lower bound $r_t\ge r\in[0,1]$, the approximation-ratio lower bound in Eq.~\eqref{eq:product_bound} can be written as
\begin{equation}\label{eq:phi_k_r}
\Phi_k(r)=1-\prod_{t=1}^{k-1}\left(1-\frac{a_k r}{t}\right),
\qquad
\eta\ge\Phi_k(r).
\end{equation}
Because each product factor satisfies $0\le 1-a_k r/t\le 1$, improving the single-split quality $r$ monotonically increases the overall approximation-ratio lower bound $\Phi_k(r)$. In particular, when every split attains the exact MaxCut of its subgraph, namely $r=1$, $\Phi_k(1)$ gives the approximation-ratio lower bound for replacing a single global $k$-cut with step-by-step binary-tree splitting.

From the single-step recurrence in Eq.~\eqref{eq:gap_recurrence}, the splitting quality $r_t$ at step $t$ acts on the difference $C_{\max}-C_t$ between the current cut value and the global optimum, rather than directly multiplying the existing cumulative cut value $C_t$. Previously cut edges are permanently retained in subsequent steps, and edges left uncut in earlier steps due to approximation errors remain inside the subgraphs $G[S]$ with opportunities to be cut in later steps. Therefore, the approximation ratio after $k-1$ splits does not decay exponentially as $r^{k-1}$. Furthermore, when $r>0$, combining $a_k>1/2$ from Eq.~\eqref{eq:ak_def} with $1-x\le e^{-x}$ and $\sum_{t=1}^{k-1}1/t>\ln k$ gives $\Phi_k(r)>1-k^{-r/2}$, showing that $\Phi_k(r)$ gradually approaches $1$ as $k$ increases.

%
\section{Numerical simulations}\label{sec:experiments}

Theoretical analysis has provided a worst-case lower bound on the approximation ratio of BT-QAOA. However, because both the outer splitting process and the inner node QAOA are approximate, the actual performance of the algorithm on specific instances requires further verification through numerical simulations. In this section, we analyze the algorithm from four aspects: approximation accuracy, scalability, propagation of approximation errors, and hardware resource overhead.

\subsection{Structural loss under exact bisection}\label{subsec:exp_exact_bisection}

BT-QAOA transforms the Max-$k$-Cut problem into a sequence of standard MaxCut subproblems, and whether this transformation can closely approach the global optimum directly determines the solution quality of the overall algorithm. To this end, we assume that each subproblem is solved as an exact MaxCut, which isolates the effect of the outer splitting process from node-solver errors. Denote the approximation ratio in this setting by $\eta_{\mathrm{exact}}$. The methods for finding the node MaxCut values and the full-graph optimum $C_{\max}$ are given in Appendix~\ref{app:subsec_exact_solver}.

\begin{figure}[t]
\centering
\includegraphics[width=\columnwidth]{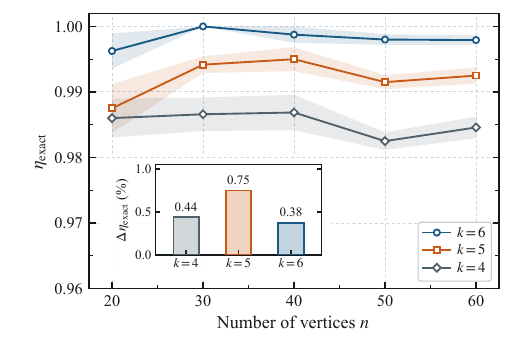}
\caption{Approximation ratio $\eta_{\mathrm{exact}}$ versus the number of vertices $n$ for $k=4,5,6$ on unweighted random $8$-regular graphs. Each data point is the mean over $10$ independent graph instances, and the light shaded bands indicate one standard error above and below the mean. The inset shows the fluctuation range $\Delta\eta_{\mathrm{exact}}$ of each curve across different values of $n$.}\label{fig:exact_scaling}
\end{figure}

As seen from Fig.~\ref{fig:exact_scaling}, for $k=4,5,6$, the mean $\eta_{\mathrm{exact}}$ across different vertex counts $n$ remains no less than $0.9825$, $0.9875$, and $0.9963$, respectively. As $n$ increases, the fluctuation range $\Delta\eta_{\mathrm{exact}}$ of each of the three curves stays below $0.8\%$, indicating no clear correlation between $\eta_{\mathrm{exact}}$ and graph size. Meanwhile, at a fixed graph size, $\eta_{\mathrm{exact}}$ increases successively with $k$. From the perspective of sequential splitting, the algorithm divides only a single subset into two at each step, leaving inside the subsets during early steps some edges that could be cut under a multiway partition. As $k$ increases and more splits are performed, these edges have more opportunities to be cut, bringing the final result closer to the global optimum.

\subsection{Splitting quality and error propagation}\label{subsec:exp_quality_propagation}

In practical solving, a single split usually cannot attain the exact MaxCut of a subgraph, making it necessary to clarify how approximation errors evolve across multiple splits and how different splitting qualities affect the final accuracy. We preset a quality lower bound $q$ and require the split constructed on each subgraph $G[S]$ to satisfy $g(S)\ge qM(S)$ while staying as close to $qM(S)$ as possible. The construction method is described in Appendix~\ref{app:subsec_quality_solver}. Taking exact bisection on the same graph as the baseline, let the cumulative cut values of the two procedures after step $s$ be $C_{\mathrm{exact},s+1}$ and $C_{q,s+1}$, respectively, and define the normalized cut-value gap
\begin{equation}\label{eq:gap_def}
\delta_s=\frac{C_{\mathrm{exact},s+1}-C_{q,s+1}}{C_{\mathrm{exact}}},
\end{equation}
where the denominator $C_{\mathrm{exact}}$ is the final cut value of exact bisection.

\begin{figure}[t]
\centering
\includegraphics[width=\columnwidth]{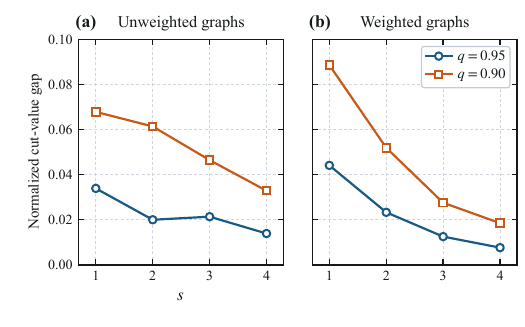}
\caption{Normalized cut-value gap $\delta_s$ versus the split step $s$ on random $10$-regular graphs with $n=60$ and $k=5$. (a) Unweighted graphs. (b) Weighted graphs with edge weights following the log-normal distribution $\ln w_e\sim\mathcal{N}(-1.125,2.25)$. Each data point is the mean over $10$ independent graph instances.}\label{fig:unweighted_quality}\label{fig:weighted_quality}
\end{figure}

From Figs.~\ref{fig:unweighted_quality}(a) and \ref{fig:weighted_quality}(b), on both unweighted graphs and weighted graphs of the same topology with continuous edge weights, the $\delta_s$ values of all curves do not accumulate and grow with the number of splits, but instead exhibit an overall decline as splitting proceeds. Specifically, at $q=0.90$, the mean $\delta_s$ on the two types of graphs drops from $0.0678$ and $0.0886$ at the first step to $0.0329$ and $0.0185$ at the fourth step. Meanwhile, the final mean approximation ratios still reach $0.9510$ and $0.9775$, respectively, exceeding the theoretical lower bound $\Phi_5(0.90)=0.7618$ given by Eq.~\eqref{eq:phi_k_r}. On the weighted graphs, the mean $g(S)/M(S)$ for $q=0.95$ and $q=0.90$ remains stable at $0.950\text{--}0.952$ and $0.900\text{--}0.901$. This shows that the decline in $\delta_s$ does not stem from higher single-split quality in later steps, but from the sequential splitting process itself. Further comparing different splitting qualities shows that across all steps on both graph types, $\delta_s$ at $q=0.95$ stays significantly below that at $q=0.90$. The final $\delta_4$ values are $0.0139$ and $0.0076$, both less than half of those in the $q=0.90$ case. This indicates that although the sequential splitting structure suppresses the accumulation of approximation errors, improving the splitting quality at individual nodes remains key to reducing the final cut-value loss.

\subsection{Approximation accuracy of BT-QAOA in the noiseless regime}\label{subsec:exp_noiseless_qaoa}

The preceding two subsections used exact MaxCut and node splits of preset quality, respectively, without yet accounting for the effects of QAOA variational training and finite sampling. To examine the solution accuracy of the complete BT-QAOA with shallow circuits, this subsection uses $p=2$ standard MaxCut QAOA to solve each node subgraph and performs noiseless statevector simulations. Detailed training and sampling settings are given in Appendix~\ref{app:subsec_qaoa_impl}.

\begin{figure}[!t]
\centering
\includegraphics[width=\columnwidth]{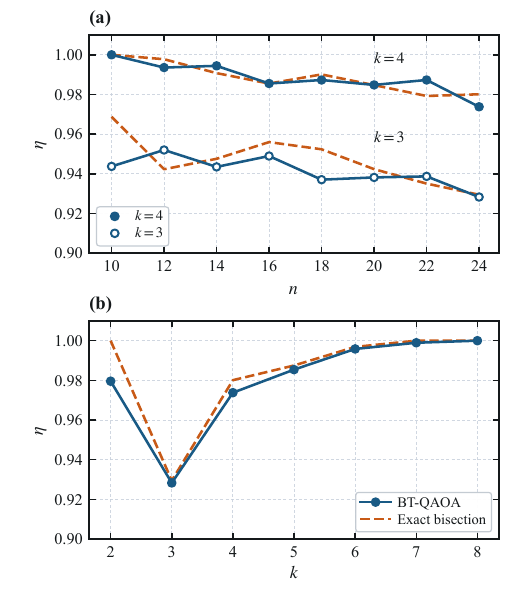}
\caption{Comparison between BT-QAOA at $p=2$ and exact bisection on unweighted random $8$-regular graphs. (a) Approximation ratio $\eta$ versus $n$ for $k=3,4$. (b) Variation of $\eta$ with $k$ at $n=24$. Each data point is the mean over $10$ independent graph instances.}\label{fig:noiseless_qaoa}
\end{figure}

From Fig.~\ref{fig:noiseless_qaoa}(a), at $p=2$, the mean $\eta$ for $k=3$ and $k=4$ across different values of $n$ stays above $0.9283$ and $0.9738$, respectively. In Fig.~\ref{fig:noiseless_qaoa}(b), for $k\ge 4$, the mean $\eta$ gradually approaches $1$ as $k$ increases. Compared with exact bisection, all data points in the figure satisfy $|\eta-\eta_{\mathrm{exact}}|\le 0.0250$, showing that replacing the exact solver with shallow QAOA under finite sampling still achieves solution quality close to that of exact bisection. Notably, at $k=3$, the mean $\eta$ of BT-QAOA and exact bisection in Fig.~\ref{fig:noiseless_qaoa}(b) drop together to $0.9283$ and $0.9295$. Their close agreement indicates that this drop is mainly tied to the outer sequential splitting structure. Once the first step fixes a partition under the MaxCut objective, the second step can only continue splitting one of the two subsets and cannot readjust vertex assignments between the two sides from the first step. Therefore, a locally optimal MaxCut need not correspond to a globally optimal three-way cut. On the same $n=10,12$ instances (requiring $20$ and $24$ qubits under binary encoding, whereas $n\ge 14$ requires at least $28$ qubits), we further compare BT-QAOA with $p=2$ binary-encoded QAOA allocated $2k-3$ times the single-node sampling budget. At $k=4$, both methods reach $\eta=1.0000$ at $n=10$, and BT-QAOA achieves $0.9936$ versus $0.9957$ at $n=12$, differing by only $0.0021$ while using half as many qubits. At $k=3$, BT-QAOA obtains $0.9438$ and $0.9520$ at $n=10$ and $12$, compared with $1.0000$ and $0.9904$ for binary encoding. The resource advantage of BT-QAOA becomes increasingly prominent as $k$ grows, whereas for small- and medium-sized Max-$3$-Cut instances within the hardware budget, one-hot or binary encoding is also a viable alternative.

\subsection{Compiled resource overhead and estimated circuit success probability}\label{subsec:exp_noisy_qaoa}

Section~\ref{sec:resources} compared logical resources under all-to-all connectivity, whereas on actual quantum processors, connectivity constraints among physical qubits and operation errors likewise restrict the executable scale of circuits~\cite{Weidenfeller2022,Li2019}. One-hot encoding requires $kn$ qubits, and handling single-vertex constraints introduces additional two-qubit gate overhead. Therefore, this subsection selects the more resource-efficient $k=3,4$ binary encodings as comparison baselines. Using the coupling topology and calibration parameters of the IBM Torino processor, we compile at $p=2$ both the binary-encoded QAOA circuits and the BT-QAOA root-node circuit, which sets the single-node resource upper bound. Here, binary encoding for $k=3$ and $k=4$ adopts the low-gate-count constructions in Appendix~\ref{app:resources} with $8$ and $6$ logical CNOT gates per edge per layer, respectively. Detailed parameter settings are given in Appendix~\ref{app:subsec_qaoa_impl}. For the compiled circuits, we count the native two-qubit CZ gates and calculate the circuit estimated success probability from the calibrated error rate $\epsilon_g$ of each operation~\cite{Nishio2020},
\begin{equation}\label{eq:esp_def}
\mathrm{ESP}=\prod_{g\in\mathcal{G}_{\mathrm{compiled}}}(1-\epsilon_g),
\end{equation}
where $\mathcal{G}_{\mathrm{compiled}}$ is the set of all quantum gates and measurement operations in the circuit.

\begin{figure}[!tb]
\centering
\includegraphics[width=\columnwidth]{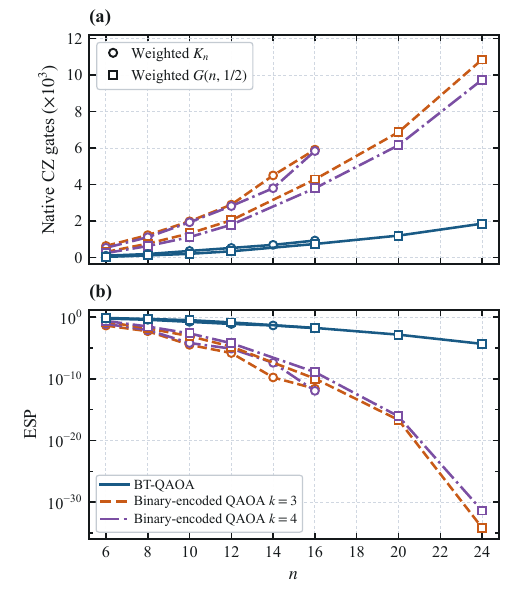}
\caption{Hardware compilation comparison between the BT-QAOA root-node circuit and $k=3,4$ binary-encoded QAOA at $p=2$ on weighted complete graphs $K_n$ and weighted Erd\H{o}s--R\'enyi random graphs $G(n,1/2)$, with edge weights drawn independently from $U(0.01,1)$. (a) Arithmetic mean of the compiled native CZ gate count. (b) Geometric mean of $\mathrm{ESP}$. Each data point is obtained from $5$ independent graph instances.}\label{fig:noisy_qaoa}
\end{figure}

From Fig.~\ref{fig:noisy_qaoa}(a), as $n$ increases, the compiled CZ gate count of binary encoding grows rapidly, whereas the gate count of the BT-QAOA root-node circuit increases much more gently. Taking $24$-vertex random graphs $G(n,1/2)$ as an example, the mean CZ gate counts of $k=3,4$ binary encodings are $10855.2$ and $9748.8$, respectively, whereas BT-QAOA reduces this count to $1866.4$. The former are about $5.8$ and $5.2$ times the latter. Because the factors $1-\epsilon_g$ in Eq.~\eqref{eq:esp_def} multiply along the circuit, the $\mathrm{ESP}$ of BT-QAOA in Fig.~\ref{fig:noisy_qaoa}(b) decays much more gently with $n$ than that of binary encoding. Moreover, it remains on a comparable order of magnitude between denser complete graphs and random graphs of the same size. Even on $16$-vertex complete graphs and $24$-vertex random graphs, its $\mathrm{ESP}$ remains $1.68\times 10^{-2}$ and $4.70\times 10^{-5}$, improving over binary encoding by approximately $10$ and more than $27$ orders of magnitude, respectively. This means that by transforming multiway partitioning into standard MaxCut circuits, BT-QAOA effectively controls the compiled gate count under a restricted coupling topology and mitigates the accumulation of operation errors within a single circuit.

%
\section{Conclusion}\label{sec:discussion}

Through step-by-step binary-tree splitting, BT-QAOA proposed in this work transforms the Max-$k$-Cut problem into standard MaxCut subproblems on at most $2k-3$ induced subgraphs. As a result, the number of qubits and the single-circuit gate count required to solve this problem no longer grow with $k$. Theoretically, we link the splitting quality on individual node subgraphs to the final partition, providing an overall approximation guarantee for the sequential splitting procedure. Numerical simulations show that given a splitting-quality lower bound, the cut-value loss does not accumulate and amplify as the number of splits increases. Moreover, in noiseless statevector simulations, node QAOA at $p=2$ achieves approximation ratios close to those of exact bisection. Under hardware compilation on a restricted coupling topology, compared with binary-encoded QAOA, BT-QAOA substantially reduces the native two-qubit gate count and improves the estimated success probability of a single circuit by orders of magnitude. These results support our proposed idea of solving the Max-$k$-Cut problem from the perspective of problem decomposition, and show that multiway partitioning problems can be effectively solved using standard MaxCut circuits alone without relying on complex encodings. This substantially expands the applicability of QAOA on near-term quantum devices.

In the future, exploring whether search strategies that retain multiple splitting results can improve the solution performance at $k=3$ is a worthwhile direction for further study. Furthermore, because each node imposes no additional constraints and consistently retains the standard MaxCut form, existing MaxCut optimization methods can be directly applied to improve the solution at each node. For example, recursive variable elimination~\cite{Bravyi2020}, warm-starting~\cite{Egger2021}, or adaptive operator construction~\cite{Zhu2022} can be incorporated to improve the splitting quality of shallow circuits and compress the gate count at the root node. Meanwhile, parameter transfer~\cite{Shaydulin2023,Galda2023} can be leveraged to reduce the repeated training overhead across multiple subproblems, thereby advancing the application of the algorithm to larger-scale practical engineering problems.

%
\appendix

\section{Encoded circuit constructions and quantum resources}\label{app:resources}

When $k=2^L$, binary encoding assigns $L=\log_2 k$ qubits to each vertex, and the same-group indicator operator in Eq.~\eqref{eq:binary_same_group} can be expanded as
\begin{equation}\label{eq:app_pauli_expansion}
\begin{aligned}
\One[c_i=c_j]
&=2^{-L}\prod_{\ell=1}^{L}(I+Z_{i\ell}Z_{j\ell}) \\
&=2^{-L}\sum_{T\subseteq\{1,\dots,L\}}\prod_{\ell\in T} Z_{i\ell}Z_{j\ell}.
\end{aligned}
\end{equation}
Here the empty set $T=\varnothing$ contributes only a global phase, while the remaining $2^L-1$ nonempty subsets correspond to distinct Pauli-$Z$ product terms. There are $\binom{L}{s}$ subsets of size $|T|=s$, each acting on $2s$ qubits and requiring $2(2s-1)$ logical CNOT gates under standard CNOT-ladder decomposition. Summing over $s=1,\dots,L$, the number of logical CNOT gates required per edge in a single cost-evolution layer is
\begin{equation}\label{eq:app_naive_cnot}
\begin{aligned}
\sum_{s=1}^{L}\binom{L}{s}2(2s-1)
&=4L\sum_{s=1}^{L}\binom{L-1}{s-1}-2\sum_{s=1}^{L}\binom{L}{s} \\
&=(L-1)2^{L+1}+2.
\end{aligned}
\end{equation}

Reusing intermediate parities further compresses the gate count. In the computational basis $|z_{i1}\cdots z_{iL}z_{j1}\cdots z_{jL}\rangle$, the eigenvalue of each term $\prod_{\ell\in T} Z_{i\ell}Z_{j\ell}$ in Eq.~\eqref{eq:app_pauli_expansion} is determined solely by the parity sum $\bigoplus_{\ell\in T} x_\ell$ of the bitwise XOR variables $x_\ell=z_{i\ell}\oplus z_{j\ell}$. One can therefore first apply CNOT gates sequentially from $(i,\ell)$ to $(j,\ell)$ for $\ell=1,\dots,L$, using $L$ CNOT gates to store $(x_1,\dots,x_L)$ in the register of vertex $j$. Next, all $2^L-1$ nonempty XOR combinations of $\{x_1,\dots,x_L\}$ are traversed on that register with the corresponding $R_Z$ gates applied term by term, before finally restoring the register to $(x_1,\dots,x_L)$. Denote the number of CNOT gates required to complete this traversal and restore the state by $q_L$; clearly $q_1=0$. For $L\ge 2$, fixing the $L$th qubit holding $x_L$ as the target and applying $2^{L-1}$ CNOT gates sequentially following the cyclic Gray-code order of the first $L-1$ qubits~\cite{Bullock2004} traverses all parity terms containing $x_L$ while restoring that qubit to $x_L$. The remaining parity terms not containing $x_L$ involve only the first $L-1$ qubits and can be completed recursively using $q_{L-1}$ CNOT gates, giving $q_L=q_{L-1}+2^{L-1}$ and hence $q_L=2^L-2$. Finally, another $L$ CNOT gates undo the initial bitwise XOR, yielding a total logical CNOT gate count of $2L+q_L=2L+2^L-2$.

When $L=2$, namely $k=4$, for nondegenerate edge phases, the above construction uses $6$ logical CNOT gates, attaining the minimum CNOT count among ancilla-free circuits composed of CNOT and single-qubit $R_Z$ gates. Single-qubit $Z$ rotations do not alter XOR relations in the computational basis~\cite{Amy2019}. However, the four-body term in Eq.~\eqref{eq:app_pauli_expansion} requires the circuit to synthesize the four-bit XOR $z_{i1}\oplus z_{j1}\oplus z_{i2}\oplus z_{j2}$ on a single qubit at some point and restore all four qubits at the end. Because each CNOT gate only adds the current XOR expressions of two qubits, starting from single-qubit initial states, the first two CNOT gates can synthesize at most two-bit and three-bit XORs. Therefore, synthesizing a four-bit XOR requires at least $3$ CNOT gates. By the reversibility of CNOT gates, restoring the state from that point back to the initial state likewise requires at least $3$ CNOT gates, so the total gate count is no less than $6$.

When $k\ne 2^L$, the single-edge cost evolution depends on how redundant encoded states are merged. In the compilation comparison for $k=3$ in Sec.~\ref{subsec:exp_noisy_qaoa}, we take $L=2$, assign $00_2,01_2,10_2$ to the three subsets, respectively, and merge $11_2$ into the third subset. The corresponding four-qubit diagonal phase operator is implemented with $8$ logical CNOT gates via Walsh expansion~\cite{Welch2014}.

\section{Supplementary notes on numerical simulations}\label{app:reproducibility}

\subsection{Exact solution of MaxCut and global Max-$k$-Cut}\label{app:subsec_exact_solver}

Both the unweighted random $8$-regular graphs in Sec.~\ref{subsec:exp_exact_bisection} and the random $10$-regular graphs in Sec.~\ref{subsec:exp_quality_propagation} are generated using NetworkX~\cite{Hagberg2008}. To find the exact MaxCut value $M(S)$ of a tree node, we formulate MaxCut on the subgraph $G[S]$ as an integer linear programming model and solve it to optimality. To find the full-graph optimum $C_{\max}$, we first check whether the original graph is $k$-colorable using the saturation-degree-first backtracking algorithm~\cite{Brelaz1979}. Whenever a valid $k$-coloring exists, all edges of the original graph can be cut simultaneously, so $C_{\max}$ equals the total edge weight of the graph. This holds for all random $8$-regular graphs at $k=5,6$ and unweighted $10$-regular graphs at $k=5$. For graph instances that are not $k$-colorable, $C_{\max}$ is solved exactly via integer linear programming.

\subsection{Construction of splits of prescribed quality and weight generation}\label{app:subsec_quality_solver}

In the weighted-graph experiments of Sec.~\ref{subsec:exp_quality_propagation}, edge weights are drawn independently from the log-normal distribution $\ln w_e\sim\mathcal{N}(-1.125,2.25)$, which has a theoretical mean of $1$. In the $n=60$ instances, the heaviest $5\%$ of edges account for $46.32\%$ of the total edge weight on average. To construct an approximate split on subgraph $G[S]$ satisfying the quality lower bound $q\in\{0.95,0.90\}$, we represent vertex assignments by a bitstring $z\in\{0,1\}^{|S|}$ that is neither all-$0$ nor all-$1$, and introduce a binary variable $y_{uv}\in\{0,1\}$ for each internal edge $\{u,v\}\in E(S)$. Imposing the linear constraints $y_{uv}\le z_u+z_v$, $y_{uv}\le 2-z_u-z_v$, $y_{uv}\ge z_u-z_v$, and $y_{uv}\ge z_v-z_u$ enforces $y_{uv}=z_u\oplus z_v$. We then minimize $g(S)$ via mixed-integer linear programming subject to $g(S)=\sum_{\{u,v\}\in E(S)} w_{uv}y_{uv}\ge qM(S)$, so that the resulting cut value stays as close to $qM(S)$ as possible.

\subsection{QAOA simulation and hardware compilation settings}\label{app:subsec_qaoa_impl}

In the noiseless statevector simulations of Sec.~\ref{subsec:exp_noiseless_qaoa}, the number of vertices is set to $n=10,12,\dots,24$, with $10$ unweighted random $8$-regular graph instances generated independently at each size, and no local-search postprocessing is applied to the output partitions. When solving a single node, variational parameters are first optimized at $p=1$ from the initial point $(\gamma_1,\beta_1)=(0.3,0.35)$ using COBYLA~\cite{Powell1994}, after which initial parameters for $p=2$ are generated via linear interpolation~\cite{Zhou2020} and further optimized. Both stages use at most $30$ objective-function evaluations, with $200$ shots per evaluation to estimate the mean cut value. After optimization, $100$ independent samples are drawn and the bitstring with the highest cut value is selected to determine the node split.

\begin{table}[!htbp]
\caption{Single-instance hardware compilation comparison for a weighted $6$-vertex complete graph at $p=2$, where two-qubit gates are device-native CZ gates.}\label{tab:k6_compilation}
\footnotesize
\setlength{\tabcolsep}{3.5pt}
\begin{ruledtabular}
\begin{tabular}{llrrrr}
Method & Circuit & Qubits & CZ gates & Depth & $\mathrm{ESP}$ \\
\colrule
BT-QAOA & Root & 6  & 110 & 263  & $0.603631$ \\
Binary encoding & $k=3$   & 12 & 646 & 988  & $0.043784$ \\
Binary encoding & $k=4$   & 12 & 542 & 737  & $0.080698$ \\
\end{tabular}
\end{ruledtabular}
\end{table}

The hardware compilation in Sec.~\ref{subsec:exp_noisy_qaoa} is performed using Qiskit~\cite{JavadiAbhari2024} based on the coupling topology and calibration data of the IBM Torino processor from February 26, 2025. The transpiler optimization level is uniformly set to $3$, and faulty couplers with calibrated two-qubit error rates of at least $0.5$ are excluded. In the experiments, weighted complete graphs $K_n$ use $n=6,8,10,12,14,16$ and weighted random graphs $G(n,1/2)$ use $n=6,8,10,12,16,20,24$, with $5$ independent graph instances at each size. Table~\ref{tab:k6_compilation} lists the single-instance compilation results for a $6$-vertex weighted complete graph at $p=2$.

\bibliography{references}

\begin{thebibliography}{40}%
\makeatletter
\providecommand \@ifxundefined [1]{%
 \@ifx{#1\undefined}
}%
\providecommand \@ifnum [1]{%
 \ifnum #1\expandafter \@firstoftwo
 \else \expandafter \@secondoftwo
 \fi
}%
\providecommand \@ifx [1]{%
 \ifx #1\expandafter \@firstoftwo
 \else \expandafter \@secondoftwo
 \fi
}%
\providecommand \natexlab [1]{#1}%
\providecommand \enquote  [1]{``#1''}%
\providecommand \bibnamefont  [1]{#1}%
\providecommand \bibfnamefont [1]{#1}%
\providecommand \citenamefont [1]{#1}%
\providecommand \href@noop [0]{\@secondoftwo}%
\providecommand \href [0]{\begingroup \@sanitize@url \@href}%
\providecommand \@href[1]{\@@startlink{#1}\@@href}%
\providecommand \@@href[1]{\endgroup#1\@@endlink}%
\providecommand \@sanitize@url [0]{\catcode `\\12\catcode `\$12\catcode
  `\&12\catcode `\#12\catcode `\^12\catcode `\_12\catcode `\%12\relax}%
\providecommand \@@startlink[1]{}%
\providecommand \@@endlink[0]{}%
\providecommand \url  [0]{\begingroup\@sanitize@url \@url }%
\providecommand \@url [1]{\endgroup\@href {#1}{\urlprefix }}%
\providecommand \urlprefix  [0]{URL }%
\providecommand \Eprint [0]{\href }%
\providecommand \doibase [0]{https://doi.org/}%
\providecommand \selectlanguage [0]{\@gobble}%
\providecommand \bibinfo  [0]{\@secondoftwo}%
\providecommand \bibfield  [0]{\@secondoftwo}%
\providecommand \translation [1]{[#1]}%
\providecommand \BibitemOpen [0]{}%
\providecommand \bibitemStop [0]{}%
\providecommand \bibitemNoStop [0]{.\EOS\space}%
\providecommand \EOS [0]{\spacefactor3000\relax}%
\providecommand \BibitemShut  [1]{\csname bibitem#1\endcsname}%
\let\auto@bib@innerbib\@empty
\bibitem [{\citenamefont {Karp}(1972)}]{Karp1972}%
  \BibitemOpen
  \bibfield  {author} {\bibinfo {author} {\bibfnamefont {R.~M.}\ \bibnamefont
  {Karp}},\ }\bibfield  {title} {\bibinfo {title} {Reducibility among
  combinatorial problems},\ }in\ \href
  {https://doi.org/10.1007/978-1-4684-2001-2_9} {\emph {\bibinfo {booktitle}
  {Complexity of Computer Computations}}}\ (\bibinfo  {publisher} {Plenum},\
  \bibinfo {address} {New York},\ \bibinfo {year} {1972})\ pp.\ \bibinfo
  {pages} {85--103}\BibitemShut {NoStop}%
\bibitem [{\citenamefont {Frieze}\ and\ \citenamefont
  {Jerrum}(1997)}]{Frieze1997}%
  \BibitemOpen
  \bibfield  {author} {\bibinfo {author} {\bibfnamefont {A.}~\bibnamefont
  {Frieze}}\ and\ \bibinfo {author} {\bibfnamefont {M.}~\bibnamefont
  {Jerrum}},\ }\bibfield  {title} {\bibinfo {title} {Improved approximation
  algorithms for {MAX} $k$-{CUT} and {MAX BISECTION}},\ }\href
  {https://doi.org/10.1007/BF02523688} {\bibfield  {journal} {\bibinfo
  {journal} {Algorithmica}\ }\textbf {\bibinfo {volume} {18}},\ \bibinfo
  {pages} {67} (\bibinfo {year} {1997})}\BibitemShut {NoStop}%
\bibitem [{\citenamefont {de~Klerk}\ \emph {et~al.}(2004)\citenamefont
  {de~Klerk}, \citenamefont {Pasechnik},\ and\ \citenamefont
  {Warners}}]{deKlerk2004}%
  \BibitemOpen
  \bibfield  {author} {\bibinfo {author} {\bibfnamefont {E.}~\bibnamefont
  {de~Klerk}}, \bibinfo {author} {\bibfnamefont {D.~V.}\ \bibnamefont
  {Pasechnik}},\ and\ \bibinfo {author} {\bibfnamefont {J.~P.}\ \bibnamefont
  {Warners}},\ }\bibfield  {title} {\bibinfo {title} {On approximate graph
  colouring and {MAX-$k$-CUT} algorithms based on the $\vartheta$-function},\
  }\href {https://doi.org/10.1023/B:JOCO.0000038911.67280.3f} {\bibfield
  {journal} {\bibinfo  {journal} {J. Comb. Optim.}\ }\textbf {\bibinfo {volume}
  {8}},\ \bibinfo {pages} {267} (\bibinfo {year} {2004})}\BibitemShut {NoStop}%
\bibitem [{\citenamefont {Eisenbl{\"a}tter}\ \emph {et~al.}(2002)\citenamefont
  {Eisenbl{\"a}tter}, \citenamefont {Gr{\"o}tschel},\ and\ \citenamefont
  {Koster}}]{Eisenblaetter2002}%
  \BibitemOpen
  \bibfield  {author} {\bibinfo {author} {\bibfnamefont {A.}~\bibnamefont
  {Eisenbl{\"a}tter}}, \bibinfo {author} {\bibfnamefont {M.}~\bibnamefont
  {Gr{\"o}tschel}},\ and\ \bibinfo {author} {\bibfnamefont {A.~M. C.~A.}\
  \bibnamefont {Koster}},\ }\bibfield  {title} {\bibinfo {title} {Frequency
  planning and ramifications of coloring},\ }\href
  {https://doi.org/10.7151/dmgt.1158} {\bibfield  {journal} {\bibinfo
  {journal} {Discuss. Math. Graph Theory}\ }\textbf {\bibinfo {volume} {22}},\
  \bibinfo {pages} {51} (\bibinfo {year} {2002})}\BibitemShut {NoStop}%
\bibitem [{\citenamefont {Carlson}\ and\ \citenamefont
  {Nemhauser}(1966)}]{Carlson1966}%
  \BibitemOpen
  \bibfield  {author} {\bibinfo {author} {\bibfnamefont {R.~C.}\ \bibnamefont
  {Carlson}}\ and\ \bibinfo {author} {\bibfnamefont {G.~L.}\ \bibnamefont
  {Nemhauser}},\ }\bibfield  {title} {\bibinfo {title} {Scheduling to minimize
  interaction cost},\ }\href {https://doi.org/10.1287/opre.14.1.52} {\bibfield
  {journal} {\bibinfo  {journal} {Oper. Res.}\ }\textbf {\bibinfo {volume}
  {14}},\ \bibinfo {pages} {52} (\bibinfo {year} {1966})}\BibitemShut {NoStop}%
\bibitem [{\citenamefont {Gaur}\ \emph {et~al.}(2008)\citenamefont {Gaur},
  \citenamefont {Krishnamurti},\ and\ \citenamefont {Kohli}}]{Gaur2008}%
  \BibitemOpen
  \bibfield  {author} {\bibinfo {author} {\bibfnamefont {D.~R.}\ \bibnamefont
  {Gaur}}, \bibinfo {author} {\bibfnamefont {R.}~\bibnamefont {Krishnamurti}},\
  and\ \bibinfo {author} {\bibfnamefont {R.}~\bibnamefont {Kohli}},\ }\bibfield
   {title} {\bibinfo {title} {The capacitated max $k$-cut problem},\ }\href
  {https://doi.org/10.1007/s10107-007-0139-z} {\bibfield  {journal} {\bibinfo
  {journal} {Math. Program.}\ }\textbf {\bibinfo {volume} {115}},\ \bibinfo
  {pages} {65} (\bibinfo {year} {2008})}\BibitemShut {NoStop}%
\bibitem [{\citenamefont {Fuchs}\ \emph {et~al.}(2021)\citenamefont {Fuchs},
  \citenamefont {Kolden}, \citenamefont {Aase},\ and\ \citenamefont
  {Sartor}}]{Fuchs2021}%
  \BibitemOpen
  \bibfield  {author} {\bibinfo {author} {\bibfnamefont {F.~G.}\ \bibnamefont
  {Fuchs}}, \bibinfo {author} {\bibfnamefont {H.~{\O}.}\ \bibnamefont
  {Kolden}}, \bibinfo {author} {\bibfnamefont {N.~H.}\ \bibnamefont {Aase}},\
  and\ \bibinfo {author} {\bibfnamefont {G.}~\bibnamefont {Sartor}},\
  }\bibfield  {title} {\bibinfo {title} {Efficient encoding of the weighted
  {MAX} $k$-{CUT} on a quantum computer using {QAOA}},\ }\href
  {https://doi.org/10.1007/s42979-020-00437-z} {\bibfield  {journal} {\bibinfo
  {journal} {SN Comput. Sci.}\ }\textbf {\bibinfo {volume} {2}},\ \bibinfo
  {pages} {89} (\bibinfo {year} {2021})}\BibitemShut {NoStop}%
\bibitem [{\citenamefont {Fuchs}\ \emph {et~al.}(2025)\citenamefont {Fuchs},
  \citenamefont {Pariente~Bassa},\ and\ \citenamefont {Lien}}]{Fuchs2025}%
  \BibitemOpen
  \bibfield  {author} {\bibinfo {author} {\bibfnamefont {F.~G.}\ \bibnamefont
  {Fuchs}}, \bibinfo {author} {\bibfnamefont {R.}~\bibnamefont
  {Pariente~Bassa}},\ and\ \bibinfo {author} {\bibfnamefont {F.}~\bibnamefont
  {Lien}},\ }\bibfield  {title} {\bibinfo {title} {Encodings of the weighted
  {MAX} $k$-{CUT} problem on qubit systems},\ }\href
  {https://doi.org/10.3389/frqst.2025.1636042} {\bibfield  {journal} {\bibinfo
  {journal} {Front. Quantum Sci. Technol.}\ }\textbf {\bibinfo {volume} {4}},\
  \bibinfo {pages} {1636042} (\bibinfo {year} {2025})}\BibitemShut {NoStop}%
\bibitem [{\citenamefont {Apte}\ \emph {et~al.}(2026)\citenamefont {Apte},
  \citenamefont {Boulebnane}, \citenamefont {Jin}, \citenamefont {Omanakuttan},
  \citenamefont {Perlin},\ and\ \citenamefont {Shaydulin}}]{Apte2026a}%
  \BibitemOpen
  \bibfield  {author} {\bibinfo {author} {\bibfnamefont {A.}~\bibnamefont
  {Apte}}, \bibinfo {author} {\bibfnamefont {S.}~\bibnamefont {Boulebnane}},
  \bibinfo {author} {\bibfnamefont {Y.}~\bibnamefont {Jin}}, \bibinfo {author}
  {\bibfnamefont {S.}~\bibnamefont {Omanakuttan}}, \bibinfo {author}
  {\bibfnamefont {M.~A.}\ \bibnamefont {Perlin}},\ and\ \bibinfo {author}
  {\bibfnamefont {R.}~\bibnamefont {Shaydulin}},\ }\bibfield  {title} {\bibinfo
  {title} {Quantum approximate optimization of integer graph problems and
  surpassing semidefinite programming for {Max-$k$-Cut}},\ }\href@noop {}
  {\bibfield  {journal} {\bibinfo  {journal} {arXiv:2602.05956}\ } (\bibinfo
  {year} {2026})}\BibitemShut {NoStop}%
\bibitem [{\citenamefont {Farhi}\ \emph {et~al.}(2014)\citenamefont {Farhi},
  \citenamefont {Goldstone},\ and\ \citenamefont {Gutmann}}]{Farhi2014}%
  \BibitemOpen
  \bibfield  {author} {\bibinfo {author} {\bibfnamefont {E.}~\bibnamefont
  {Farhi}}, \bibinfo {author} {\bibfnamefont {J.}~\bibnamefont {Goldstone}},\
  and\ \bibinfo {author} {\bibfnamefont {S.}~\bibnamefont {Gutmann}},\
  }\bibfield  {title} {\bibinfo {title} {A quantum approximate optimization
  algorithm},\ }\href@noop {} {\bibfield  {journal} {\bibinfo  {journal}
  {arXiv:1411.4028}\ } (\bibinfo {year} {2014})}\BibitemShut {NoStop}%
\bibitem [{\citenamefont {Zhou}\ \emph {et~al.}(2020)\citenamefont {Zhou},
  \citenamefont {Wang}, \citenamefont {Choi}, \citenamefont {Pichler},\ and\
  \citenamefont {Lukin}}]{Zhou2020}%
  \BibitemOpen
  \bibfield  {author} {\bibinfo {author} {\bibfnamefont {L.}~\bibnamefont
  {Zhou}}, \bibinfo {author} {\bibfnamefont {S.-T.}\ \bibnamefont {Wang}},
  \bibinfo {author} {\bibfnamefont {S.}~\bibnamefont {Choi}}, \bibinfo {author}
  {\bibfnamefont {H.}~\bibnamefont {Pichler}},\ and\ \bibinfo {author}
  {\bibfnamefont {M.~D.}\ \bibnamefont {Lukin}},\ }\bibfield  {title} {\bibinfo
  {title} {Quantum approximate optimization algorithm: Performance, mechanism,
  and implementation on near-term devices},\ }\href
  {https://doi.org/10.1103/PhysRevX.10.021067} {\bibfield  {journal} {\bibinfo
  {journal} {Phys. Rev. X}\ }\textbf {\bibinfo {volume} {10}},\ \bibinfo
  {pages} {021067} (\bibinfo {year} {2020})}\BibitemShut {NoStop}%
\bibitem [{\citenamefont {Harrigan}\ \emph {et~al.}(2021)\citenamefont
  {Harrigan}, \citenamefont {Sung}, \citenamefont {Neeley}, \citenamefont
  {Satzinger}, \citenamefont {Arute}, \citenamefont {Arya}, \citenamefont
  {Atalaya}, \citenamefont {Babbush}, \citenamefont {Bacon}, \citenamefont
  {Bardin} \emph {et~al.}}]{Harrigan2021}%
  \BibitemOpen
  \bibfield  {author} {\bibinfo {author} {\bibfnamefont {M.~P.}\ \bibnamefont
  {Harrigan}}, \bibinfo {author} {\bibfnamefont {K.~J.}\ \bibnamefont {Sung}},
  \bibinfo {author} {\bibfnamefont {M.}~\bibnamefont {Neeley}}, \bibinfo
  {author} {\bibfnamefont {K.~J.}\ \bibnamefont {Satzinger}}, \bibinfo {author}
  {\bibfnamefont {F.}~\bibnamefont {Arute}}, \bibinfo {author} {\bibfnamefont
  {K.}~\bibnamefont {Arya}}, \bibinfo {author} {\bibfnamefont {J.}~\bibnamefont
  {Atalaya}}, \bibinfo {author} {\bibfnamefont {R.}~\bibnamefont {Babbush}},
  \bibinfo {author} {\bibfnamefont {D.}~\bibnamefont {Bacon}}, \bibinfo
  {author} {\bibfnamefont {J.~C.}\ \bibnamefont {Bardin}}, \emph {et~al.},\
  }\bibfield  {title} {\bibinfo {title} {Quantum approximate optimization of
  non-planar graph problems on a planar superconducting processor},\ }\href
  {https://doi.org/10.1038/s41567-020-01105-y} {\bibfield  {journal} {\bibinfo
  {journal} {Nat. Phys.}\ }\textbf {\bibinfo {volume} {17}},\ \bibinfo {pages}
  {332} (\bibinfo {year} {2021})}\BibitemShut {NoStop}%
\bibitem [{\citenamefont {He}\ \emph {et~al.}(2025)\citenamefont {He},
  \citenamefont {Amaro}, \citenamefont {Shaydulin},\ and\ \citenamefont
  {Pistoia}}]{He2025}%
  \BibitemOpen
  \bibfield  {author} {\bibinfo {author} {\bibfnamefont {Z.}~\bibnamefont
  {He}}, \bibinfo {author} {\bibfnamefont {D.}~\bibnamefont {Amaro}}, \bibinfo
  {author} {\bibfnamefont {R.}~\bibnamefont {Shaydulin}},\ and\ \bibinfo
  {author} {\bibfnamefont {M.}~\bibnamefont {Pistoia}},\ }\bibfield  {title}
  {\bibinfo {title} {Performance of quantum approximate optimization with
  quantum error detection},\ }\href
  {https://doi.org/10.1038/s42005-025-02136-8} {\bibfield  {journal} {\bibinfo
  {journal} {Commun. Phys.}\ }\textbf {\bibinfo {volume} {8}},\ \bibinfo
  {pages} {217} (\bibinfo {year} {2025})}\BibitemShut {NoStop}%
\bibitem [{\citenamefont {Wang}\ \emph {et~al.}(2018)\citenamefont {Wang},
  \citenamefont {Hadfield}, \citenamefont {Jiang},\ and\ \citenamefont
  {Rieffel}}]{Wang2018}%
  \BibitemOpen
  \bibfield  {author} {\bibinfo {author} {\bibfnamefont {Z.}~\bibnamefont
  {Wang}}, \bibinfo {author} {\bibfnamefont {S.}~\bibnamefont {Hadfield}},
  \bibinfo {author} {\bibfnamefont {Z.}~\bibnamefont {Jiang}},\ and\ \bibinfo
  {author} {\bibfnamefont {E.~G.}\ \bibnamefont {Rieffel}},\ }\bibfield
  {title} {\bibinfo {title} {Quantum approximate optimization algorithm for
  {MaxCut}: A fermionic view},\ }\href
  {https://doi.org/10.1103/PhysRevA.97.022304} {\bibfield  {journal} {\bibinfo
  {journal} {Phys. Rev. A}\ }\textbf {\bibinfo {volume} {97}},\ \bibinfo
  {pages} {022304} (\bibinfo {year} {2018})}\BibitemShut {NoStop}%
\bibitem [{\citenamefont {Wurtz}\ and\ \citenamefont {Love}(2021)}]{Wurtz2021}%
  \BibitemOpen
  \bibfield  {author} {\bibinfo {author} {\bibfnamefont {J.}~\bibnamefont
  {Wurtz}}\ and\ \bibinfo {author} {\bibfnamefont {P.~J.}\ \bibnamefont
  {Love}},\ }\bibfield  {title} {\bibinfo {title} {{MaxCut} quantum approximate
  optimization algorithm performance guarantees for $p>1$},\ }\href
  {https://doi.org/10.1103/PhysRevA.103.042612} {\bibfield  {journal} {\bibinfo
   {journal} {Phys. Rev. A}\ }\textbf {\bibinfo {volume} {103}},\ \bibinfo
  {pages} {042612} (\bibinfo {year} {2021})}\BibitemShut {NoStop}%
\bibitem [{\citenamefont {Lucas}(2014)}]{Lucas2014}%
  \BibitemOpen
  \bibfield  {author} {\bibinfo {author} {\bibfnamefont {A.}~\bibnamefont
  {Lucas}},\ }\bibfield  {title} {\bibinfo {title} {Ising formulations of many
  {NP} problems},\ }\href {https://doi.org/10.3389/fphy.2014.00005} {\bibfield
  {journal} {\bibinfo  {journal} {Front. Phys.}\ }\textbf {\bibinfo {volume}
  {2}},\ \bibinfo {pages} {5} (\bibinfo {year} {2014})}\BibitemShut {NoStop}%
\bibitem [{\citenamefont {Hadfield}\ \emph {et~al.}(2019)\citenamefont
  {Hadfield}, \citenamefont {Wang}, \citenamefont {O'Gorman}, \citenamefont
  {Rieffel}, \citenamefont {Venturelli},\ and\ \citenamefont
  {Biswas}}]{Hadfield2019}%
  \BibitemOpen
  \bibfield  {author} {\bibinfo {author} {\bibfnamefont {S.}~\bibnamefont
  {Hadfield}}, \bibinfo {author} {\bibfnamefont {Z.}~\bibnamefont {Wang}},
  \bibinfo {author} {\bibfnamefont {B.}~\bibnamefont {O'Gorman}}, \bibinfo
  {author} {\bibfnamefont {E.~G.}\ \bibnamefont {Rieffel}}, \bibinfo {author}
  {\bibfnamefont {D.}~\bibnamefont {Venturelli}},\ and\ \bibinfo {author}
  {\bibfnamefont {R.}~\bibnamefont {Biswas}},\ }\bibfield  {title} {\bibinfo
  {title} {From the quantum approximate optimization algorithm to a quantum
  alternating operator ansatz},\ }\href {https://doi.org/10.3390/a12020034}
  {\bibfield  {journal} {\bibinfo  {journal} {Algorithms}\ }\textbf {\bibinfo
  {volume} {12}},\ \bibinfo {pages} {34} (\bibinfo {year} {2019})}\BibitemShut
  {NoStop}%
\bibitem [{\citenamefont {Wang}\ \emph {et~al.}(2020)\citenamefont {Wang},
  \citenamefont {Rubin}, \citenamefont {Dominy},\ and\ \citenamefont
  {Rieffel}}]{WangXY2020}%
  \BibitemOpen
  \bibfield  {author} {\bibinfo {author} {\bibfnamefont {Z.}~\bibnamefont
  {Wang}}, \bibinfo {author} {\bibfnamefont {N.~C.}\ \bibnamefont {Rubin}},
  \bibinfo {author} {\bibfnamefont {J.~M.}\ \bibnamefont {Dominy}},\ and\
  \bibinfo {author} {\bibfnamefont {E.~G.}\ \bibnamefont {Rieffel}},\
  }\bibfield  {title} {\bibinfo {title} {{XY} mixers: Analytical and numerical
  results for the quantum alternating operator ansatz},\ }\href
  {https://doi.org/10.1103/PhysRevA.101.012320} {\bibfield  {journal} {\bibinfo
   {journal} {Phys. Rev. A}\ }\textbf {\bibinfo {volume} {101}},\ \bibinfo
  {pages} {012320} (\bibinfo {year} {2020})}\BibitemShut {NoStop}%
\bibitem [{\citenamefont {Harkness}\ \emph {et~al.}(2025)\citenamefont
  {Harkness}, \citenamefont {Validi}, \citenamefont {Fakhimi}, \citenamefont
  {Hicks}, \citenamefont {Stein}, \citenamefont {Terlaky},\ and\ \citenamefont
  {Zuluaga}}]{Harkness2025}%
  \BibitemOpen
  \bibfield  {author} {\bibinfo {author} {\bibfnamefont {A.}~\bibnamefont
  {Harkness}}, \bibinfo {author} {\bibfnamefont {H.}~\bibnamefont {Validi}},
  \bibinfo {author} {\bibfnamefont {R.}~\bibnamefont {Fakhimi}}, \bibinfo
  {author} {\bibfnamefont {I.~V.}\ \bibnamefont {Hicks}}, \bibinfo {author}
  {\bibfnamefont {S.}~\bibnamefont {Stein}}, \bibinfo {author} {\bibfnamefont
  {T.}~\bibnamefont {Terlaky}},\ and\ \bibinfo {author} {\bibfnamefont {L.~F.}\
  \bibnamefont {Zuluaga}},\ }\bibfield  {title} {\bibinfo {title}
  {Characterizing {QUBO} reformulations of the {Max-$k$-Cut} problem for
  quantum computing},\ }\href@noop {} {\bibfield  {journal} {\bibinfo
  {journal} {arXiv:2511.01108}\ } (\bibinfo {year} {2025})}\BibitemShut
  {NoStop}%
\bibitem [{\citenamefont {Deller}\ \emph {et~al.}(2023)\citenamefont {Deller},
  \citenamefont {Schmitt}, \citenamefont {Lewenstein}, \citenamefont {Lenk},
  \citenamefont {Federer}, \citenamefont {Jendrzejewski}, \citenamefont
  {Hauke},\ and\ \citenamefont {Kasper}}]{Deller2023}%
  \BibitemOpen
  \bibfield  {author} {\bibinfo {author} {\bibfnamefont {Y.}~\bibnamefont
  {Deller}}, \bibinfo {author} {\bibfnamefont {S.}~\bibnamefont {Schmitt}},
  \bibinfo {author} {\bibfnamefont {M.}~\bibnamefont {Lewenstein}}, \bibinfo
  {author} {\bibfnamefont {S.}~\bibnamefont {Lenk}}, \bibinfo {author}
  {\bibfnamefont {M.}~\bibnamefont {Federer}}, \bibinfo {author} {\bibfnamefont
  {F.}~\bibnamefont {Jendrzejewski}}, \bibinfo {author} {\bibfnamefont
  {P.}~\bibnamefont {Hauke}},\ and\ \bibinfo {author} {\bibfnamefont
  {V.}~\bibnamefont {Kasper}},\ }\bibfield  {title} {\bibinfo {title} {Quantum
  approximate optimization algorithm for qudit systems},\ }\href
  {https://doi.org/10.1103/PhysRevA.107.062410} {\bibfield  {journal} {\bibinfo
   {journal} {Phys. Rev. A}\ }\textbf {\bibinfo {volume} {107}},\ \bibinfo
  {pages} {062410} (\bibinfo {year} {2023})}\BibitemShut {NoStop}%
\bibitem [{\citenamefont {Bravyi}\ \emph {et~al.}(2022)\citenamefont {Bravyi},
  \citenamefont {Kliesch}, \citenamefont {Koenig},\ and\ \citenamefont
  {Tang}}]{Bravyi2022}%
  \BibitemOpen
  \bibfield  {author} {\bibinfo {author} {\bibfnamefont {S.}~\bibnamefont
  {Bravyi}}, \bibinfo {author} {\bibfnamefont {A.}~\bibnamefont {Kliesch}},
  \bibinfo {author} {\bibfnamefont {R.}~\bibnamefont {Koenig}},\ and\ \bibinfo
  {author} {\bibfnamefont {E.}~\bibnamefont {Tang}},\ }\bibfield  {title}
  {\bibinfo {title} {Hybrid quantum-classical algorithms for approximate graph
  coloring},\ }\href {https://doi.org/10.22331/q-2022-03-30-678} {\bibfield
  {journal} {\bibinfo  {journal} {Quantum}\ }\textbf {\bibinfo {volume} {6}},\
  \bibinfo {pages} {678} (\bibinfo {year} {2022})}\BibitemShut {NoStop}%
\bibitem [{\citenamefont {Ringbauer}\ \emph {et~al.}(2022)\citenamefont
  {Ringbauer}, \citenamefont {Meth}, \citenamefont {Postler}, \citenamefont
  {Stricker}, \citenamefont {Blatt}, \citenamefont {Schindler},\ and\
  \citenamefont {Monz}}]{Ringbauer2022}%
  \BibitemOpen
  \bibfield  {author} {\bibinfo {author} {\bibfnamefont {M.}~\bibnamefont
  {Ringbauer}}, \bibinfo {author} {\bibfnamefont {M.}~\bibnamefont {Meth}},
  \bibinfo {author} {\bibfnamefont {L.}~\bibnamefont {Postler}}, \bibinfo
  {author} {\bibfnamefont {R.}~\bibnamefont {Stricker}}, \bibinfo {author}
  {\bibfnamefont {R.}~\bibnamefont {Blatt}}, \bibinfo {author} {\bibfnamefont
  {P.}~\bibnamefont {Schindler}},\ and\ \bibinfo {author} {\bibfnamefont
  {T.}~\bibnamefont {Monz}},\ }\bibfield  {title} {\bibinfo {title} {A
  universal qudit quantum processor with trapped ions},\ }\href
  {https://doi.org/10.1038/s41567-022-01658-0} {\bibfield  {journal} {\bibinfo
  {journal} {Nat. Phys.}\ }\textbf {\bibinfo {volume} {18}},\ \bibinfo {pages}
  {1053} (\bibinfo {year} {2022})}\BibitemShut {NoStop}%
\bibitem [{\citenamefont {Blekos}\ \emph {et~al.}(2024)\citenamefont {Blekos},
  \citenamefont {Brand}, \citenamefont {Ceschini}, \citenamefont {Chou},
  \citenamefont {Li}, \citenamefont {Pandya},\ and\ \citenamefont
  {Summer}}]{Blekos2024}%
  \BibitemOpen
  \bibfield  {author} {\bibinfo {author} {\bibfnamefont {K.}~\bibnamefont
  {Blekos}}, \bibinfo {author} {\bibfnamefont {D.}~\bibnamefont {Brand}},
  \bibinfo {author} {\bibfnamefont {A.}~\bibnamefont {Ceschini}}, \bibinfo
  {author} {\bibfnamefont {C.-H.}\ \bibnamefont {Chou}}, \bibinfo {author}
  {\bibfnamefont {R.-H.}\ \bibnamefont {Li}}, \bibinfo {author} {\bibfnamefont
  {K.}~\bibnamefont {Pandya}},\ and\ \bibinfo {author} {\bibfnamefont
  {A.}~\bibnamefont {Summer}},\ }\bibfield  {title} {\bibinfo {title} {A review
  on {Quantum Approximate Optimization Algorithm} and its variants},\ }\href
  {https://doi.org/10.1016/j.physrep.2024.03.002} {\bibfield  {journal}
  {\bibinfo  {journal} {Phys. Rep.}\ }\textbf {\bibinfo {volume} {1068}},\
  \bibinfo {pages} {1} (\bibinfo {year} {2024})}\BibitemShut {NoStop}%
\bibitem [{\citenamefont {{\v{S}}{\'a}mal}(2009)}]{Samal2009}%
  \BibitemOpen
  \bibfield  {author} {\bibinfo {author} {\bibfnamefont {R.}~\bibnamefont
  {{\v{S}}{\'a}mal}},\ }\href {https://iti.mff.cuni.cz/series/2009/478.pdf}
  {\emph {\bibinfo {title} {Cubical coloring --- fractional covering by
  cuts}}},\ \bibinfo {type} {Tech. Rep.}\ \bibinfo {number} {2009-478}\
  (\bibinfo  {institution} {Institute for Theoretical Computer Science (ITI),
  Charles University},\ \bibinfo {year} {2009})\BibitemShut {NoStop}%
\bibitem [{\citenamefont {Nemhauser}\ \emph {et~al.}(1978)\citenamefont
  {Nemhauser}, \citenamefont {Wolsey},\ and\ \citenamefont
  {Fisher}}]{Nemhauser1978}%
  \BibitemOpen
  \bibfield  {author} {\bibinfo {author} {\bibfnamefont {G.~L.}\ \bibnamefont
  {Nemhauser}}, \bibinfo {author} {\bibfnamefont {L.~A.}\ \bibnamefont
  {Wolsey}},\ and\ \bibinfo {author} {\bibfnamefont {M.~L.}\ \bibnamefont
  {Fisher}},\ }\bibfield  {title} {\bibinfo {title} {An analysis of
  approximations for maximizing submodular set functions---{I}},\ }\href
  {https://doi.org/10.1007/BF01588971} {\bibfield  {journal} {\bibinfo
  {journal} {Math. Program.}\ }\textbf {\bibinfo {volume} {14}},\ \bibinfo
  {pages} {265} (\bibinfo {year} {1978})}\BibitemShut {NoStop}%
\bibitem [{\citenamefont {Weidenfeller}\ \emph {et~al.}(2022)\citenamefont
  {Weidenfeller}, \citenamefont {Valor}, \citenamefont {Gacon}, \citenamefont
  {Tornow}, \citenamefont {Bello}, \citenamefont {Woerner},\ and\ \citenamefont
  {Egger}}]{Weidenfeller2022}%
  \BibitemOpen
  \bibfield  {author} {\bibinfo {author} {\bibfnamefont {J.}~\bibnamefont
  {Weidenfeller}}, \bibinfo {author} {\bibfnamefont {L.~C.}\ \bibnamefont
  {Valor}}, \bibinfo {author} {\bibfnamefont {J.}~\bibnamefont {Gacon}},
  \bibinfo {author} {\bibfnamefont {C.}~\bibnamefont {Tornow}}, \bibinfo
  {author} {\bibfnamefont {L.}~\bibnamefont {Bello}}, \bibinfo {author}
  {\bibfnamefont {S.}~\bibnamefont {Woerner}},\ and\ \bibinfo {author}
  {\bibfnamefont {D.~J.}\ \bibnamefont {Egger}},\ }\bibfield  {title} {\bibinfo
  {title} {Scaling of the quantum approximate optimization algorithm on
  superconducting qubit based hardware},\ }\href
  {https://doi.org/10.22331/q-2022-12-07-870} {\bibfield  {journal} {\bibinfo
  {journal} {Quantum}\ }\textbf {\bibinfo {volume} {6}},\ \bibinfo {pages}
  {870} (\bibinfo {year} {2022})}\BibitemShut {NoStop}%
\bibitem [{\citenamefont {Li}\ \emph {et~al.}(2019)\citenamefont {Li},
  \citenamefont {Ding},\ and\ \citenamefont {Xie}}]{Li2019}%
  \BibitemOpen
  \bibfield  {author} {\bibinfo {author} {\bibfnamefont {G.}~\bibnamefont
  {Li}}, \bibinfo {author} {\bibfnamefont {Y.}~\bibnamefont {Ding}},\ and\
  \bibinfo {author} {\bibfnamefont {Y.}~\bibnamefont {Xie}},\ }\bibfield
  {title} {\bibinfo {title} {Tackling the qubit mapping problem for {NISQ}-era
  quantum devices},\ }in\ \href {https://doi.org/10.1145/3297858.3304023}
  {\emph {\bibinfo {booktitle} {Proceedings of the Twenty-Fourth International
  Conference on Architectural Support for Programming Languages and Operating
  Systems}}}\ (\bibinfo  {publisher} {ACM},\ \bibinfo {address} {New York},\
  \bibinfo {year} {2019})\ pp.\ \bibinfo {pages} {1001--1014}\BibitemShut
  {NoStop}%
\bibitem [{\citenamefont {Nishio}\ \emph {et~al.}(2020)\citenamefont {Nishio},
  \citenamefont {Pan}, \citenamefont {Satoh}, \citenamefont {Amano},\ and\
  \citenamefont {Van~Meter}}]{Nishio2020}%
  \BibitemOpen
  \bibfield  {author} {\bibinfo {author} {\bibfnamefont {S.}~\bibnamefont
  {Nishio}}, \bibinfo {author} {\bibfnamefont {Y.}~\bibnamefont {Pan}},
  \bibinfo {author} {\bibfnamefont {T.}~\bibnamefont {Satoh}}, \bibinfo
  {author} {\bibfnamefont {H.}~\bibnamefont {Amano}},\ and\ \bibinfo {author}
  {\bibfnamefont {R.}~\bibnamefont {Van~Meter}},\ }\bibfield  {title} {\bibinfo
  {title} {Extracting success from {IBM}'s 20-qubit machines using error-aware
  compilation},\ }\href {https://doi.org/10.1145/3386162} {\bibfield  {journal}
  {\bibinfo  {journal} {ACM J. Emerg. Technol. Comput. Syst.}\ }\textbf
  {\bibinfo {volume} {16}},\ \bibinfo {pages} {32} (\bibinfo {year}
  {2020})}\BibitemShut {NoStop}%
\bibitem [{\citenamefont {Bravyi}\ \emph {et~al.}(2020)\citenamefont {Bravyi},
  \citenamefont {Kliesch}, \citenamefont {Koenig},\ and\ \citenamefont
  {Tang}}]{Bravyi2020}%
  \BibitemOpen
  \bibfield  {author} {\bibinfo {author} {\bibfnamefont {S.}~\bibnamefont
  {Bravyi}}, \bibinfo {author} {\bibfnamefont {A.}~\bibnamefont {Kliesch}},
  \bibinfo {author} {\bibfnamefont {R.}~\bibnamefont {Koenig}},\ and\ \bibinfo
  {author} {\bibfnamefont {E.}~\bibnamefont {Tang}},\ }\bibfield  {title}
  {\bibinfo {title} {Obstacles to variational quantum optimization from
  symmetry protection},\ }\href
  {https://doi.org/10.1103/PhysRevLett.125.260505} {\bibfield  {journal}
  {\bibinfo  {journal} {Phys. Rev. Lett.}\ }\textbf {\bibinfo {volume} {125}},\
  \bibinfo {pages} {260505} (\bibinfo {year} {2020})}\BibitemShut {NoStop}%
\bibitem [{\citenamefont {Egger}\ \emph {et~al.}(2021)\citenamefont {Egger},
  \citenamefont {Mare{\v{c}}ek},\ and\ \citenamefont {Woerner}}]{Egger2021}%
  \BibitemOpen
  \bibfield  {author} {\bibinfo {author} {\bibfnamefont {D.~J.}\ \bibnamefont
  {Egger}}, \bibinfo {author} {\bibfnamefont {J.}~\bibnamefont
  {Mare{\v{c}}ek}},\ and\ \bibinfo {author} {\bibfnamefont {S.}~\bibnamefont
  {Woerner}},\ }\bibfield  {title} {\bibinfo {title} {Warm-starting quantum
  optimization},\ }\href {https://doi.org/10.22331/q-2021-06-17-479} {\bibfield
   {journal} {\bibinfo  {journal} {Quantum}\ }\textbf {\bibinfo {volume} {5}},\
  \bibinfo {pages} {479} (\bibinfo {year} {2021})}\BibitemShut {NoStop}%
\bibitem [{\citenamefont {Zhu}\ \emph {et~al.}(2022)\citenamefont {Zhu},
  \citenamefont {Tang}, \citenamefont {Barron}, \citenamefont
  {Calderon-Vargas}, \citenamefont {Mayhall}, \citenamefont {Barnes},\ and\
  \citenamefont {Economou}}]{Zhu2022}%
  \BibitemOpen
  \bibfield  {author} {\bibinfo {author} {\bibfnamefont {L.}~\bibnamefont
  {Zhu}}, \bibinfo {author} {\bibfnamefont {H.~L.}\ \bibnamefont {Tang}},
  \bibinfo {author} {\bibfnamefont {G.~S.}\ \bibnamefont {Barron}}, \bibinfo
  {author} {\bibfnamefont {F.~A.}\ \bibnamefont {Calderon-Vargas}}, \bibinfo
  {author} {\bibfnamefont {N.~J.}\ \bibnamefont {Mayhall}}, \bibinfo {author}
  {\bibfnamefont {E.}~\bibnamefont {Barnes}},\ and\ \bibinfo {author}
  {\bibfnamefont {S.~E.}\ \bibnamefont {Economou}},\ }\bibfield  {title}
  {\bibinfo {title} {Adaptive quantum approximate optimization algorithm for
  solving combinatorial problems on a quantum computer},\ }\href
  {https://doi.org/10.1103/PhysRevResearch.4.033029} {\bibfield  {journal}
  {\bibinfo  {journal} {Phys. Rev. Research}\ }\textbf {\bibinfo {volume}
  {4}},\ \bibinfo {pages} {033029} (\bibinfo {year} {2022})}\BibitemShut
  {NoStop}%
\bibitem [{\citenamefont {Shaydulin}\ \emph {et~al.}(2023)\citenamefont
  {Shaydulin}, \citenamefont {Lotshaw}, \citenamefont {Larson}, \citenamefont
  {Ostrowski},\ and\ \citenamefont {Humble}}]{Shaydulin2023}%
  \BibitemOpen
  \bibfield  {author} {\bibinfo {author} {\bibfnamefont {R.}~\bibnamefont
  {Shaydulin}}, \bibinfo {author} {\bibfnamefont {P.~C.}\ \bibnamefont
  {Lotshaw}}, \bibinfo {author} {\bibfnamefont {J.}~\bibnamefont {Larson}},
  \bibinfo {author} {\bibfnamefont {J.}~\bibnamefont {Ostrowski}},\ and\
  \bibinfo {author} {\bibfnamefont {T.~S.}\ \bibnamefont {Humble}},\ }\bibfield
   {title} {\bibinfo {title} {Parameter transfer for quantum approximate
  optimization of weighted {MaxCut}},\ }\href {https://doi.org/10.1145/3584706}
  {\bibfield  {journal} {\bibinfo  {journal} {ACM Trans. Quantum Comput.}\
  }\textbf {\bibinfo {volume} {4}},\ \bibinfo {pages} {19} (\bibinfo {year}
  {2023})}\BibitemShut {NoStop}%
\bibitem [{\citenamefont {Galda}\ \emph {et~al.}(2023)\citenamefont {Galda},
  \citenamefont {Gupta}, \citenamefont {Falla}, \citenamefont {Liu},
  \citenamefont {Lykov}, \citenamefont {Alexeev},\ and\ \citenamefont
  {Safro}}]{Galda2023}%
  \BibitemOpen
  \bibfield  {author} {\bibinfo {author} {\bibfnamefont {A.}~\bibnamefont
  {Galda}}, \bibinfo {author} {\bibfnamefont {E.}~\bibnamefont {Gupta}},
  \bibinfo {author} {\bibfnamefont {J.}~\bibnamefont {Falla}}, \bibinfo
  {author} {\bibfnamefont {X.}~\bibnamefont {Liu}}, \bibinfo {author}
  {\bibfnamefont {D.}~\bibnamefont {Lykov}}, \bibinfo {author} {\bibfnamefont
  {Y.}~\bibnamefont {Alexeev}},\ and\ \bibinfo {author} {\bibfnamefont
  {I.}~\bibnamefont {Safro}},\ }\bibfield  {title} {\bibinfo {title}
  {Similarity-based parameter transferability in the quantum approximate
  optimization algorithm},\ }\href {https://doi.org/10.3389/frqst.2023.1200975}
  {\bibfield  {journal} {\bibinfo  {journal} {Front. Quantum Sci. Technol.}\
  }\textbf {\bibinfo {volume} {2}},\ \bibinfo {pages} {1200975} (\bibinfo
  {year} {2023})}\BibitemShut {NoStop}%
\bibitem [{\citenamefont {Bullock}\ and\ \citenamefont
  {Markov}(2004)}]{Bullock2004}%
  \BibitemOpen
  \bibfield  {author} {\bibinfo {author} {\bibfnamefont {S.~S.}\ \bibnamefont
  {Bullock}}\ and\ \bibinfo {author} {\bibfnamefont {I.~L.}\ \bibnamefont
  {Markov}},\ }\bibfield  {title} {\bibinfo {title} {Asymptotically optimal
  circuits for arbitrary $n$-qubit diagonal computations},\ }\href
  {https://doi.org/10.26421/QIC4.1-3} {\bibfield  {journal} {\bibinfo
  {journal} {Quantum Inf. Comput.}\ }\textbf {\bibinfo {volume} {4}},\ \bibinfo
  {pages} {27} (\bibinfo {year} {2004})}\BibitemShut {NoStop}%
\bibitem [{\citenamefont {Amy}\ \emph {et~al.}(2019)\citenamefont {Amy},
  \citenamefont {Azimzadeh},\ and\ \citenamefont {Mosca}}]{Amy2019}%
  \BibitemOpen
  \bibfield  {author} {\bibinfo {author} {\bibfnamefont {M.}~\bibnamefont
  {Amy}}, \bibinfo {author} {\bibfnamefont {P.}~\bibnamefont {Azimzadeh}},\
  and\ \bibinfo {author} {\bibfnamefont {M.}~\bibnamefont {Mosca}},\ }\bibfield
   {title} {\bibinfo {title} {On the controlled-{NOT} complexity of
  controlled-{NOT}--phase circuits},\ }\href
  {https://doi.org/10.1088/2058-9565/aad8ca} {\bibfield  {journal} {\bibinfo
  {journal} {Quantum Sci. Technol.}\ }\textbf {\bibinfo {volume} {4}},\
  \bibinfo {pages} {015002} (\bibinfo {year} {2019})}\BibitemShut {NoStop}%
\bibitem [{\citenamefont {Welch}\ \emph {et~al.}(2014)\citenamefont {Welch},
  \citenamefont {Greenbaum}, \citenamefont {Mostame},\ and\ \citenamefont
  {Aspuru-Guzik}}]{Welch2014}%
  \BibitemOpen
  \bibfield  {author} {\bibinfo {author} {\bibfnamefont {J.}~\bibnamefont
  {Welch}}, \bibinfo {author} {\bibfnamefont {D.}~\bibnamefont {Greenbaum}},
  \bibinfo {author} {\bibfnamefont {S.}~\bibnamefont {Mostame}},\ and\ \bibinfo
  {author} {\bibfnamefont {A.}~\bibnamefont {Aspuru-Guzik}},\ }\bibfield
  {title} {\bibinfo {title} {Efficient quantum circuits for diagonal unitaries
  without ancillas},\ }\href {https://doi.org/10.1088/1367-2630/16/3/033040}
  {\bibfield  {journal} {\bibinfo  {journal} {New J. Phys.}\ }\textbf {\bibinfo
  {volume} {16}},\ \bibinfo {pages} {033040} (\bibinfo {year}
  {2014})}\BibitemShut {NoStop}%
\bibitem [{\citenamefont {Hagberg}\ \emph {et~al.}(2008)\citenamefont
  {Hagberg}, \citenamefont {Schult},\ and\ \citenamefont
  {Swart}}]{Hagberg2008}%
  \BibitemOpen
  \bibfield  {author} {\bibinfo {author} {\bibfnamefont {A.~A.}\ \bibnamefont
  {Hagberg}}, \bibinfo {author} {\bibfnamefont {D.~A.}\ \bibnamefont
  {Schult}},\ and\ \bibinfo {author} {\bibfnamefont {P.~J.}\ \bibnamefont
  {Swart}},\ }\bibfield  {title} {\bibinfo {title} {Exploring network
  structure, dynamics, and function using {NetworkX}},\ }in\ \href@noop {}
  {\emph {\bibinfo {booktitle} {Proceedings of the 7th Python in Science
  Conference}}},\ \bibinfo {editor} {edited by\ \bibinfo {editor}
  {\bibfnamefont {G.}~\bibnamefont {Varoquaux}}, \bibinfo {editor}
  {\bibfnamefont {T.}~\bibnamefont {Vaught}},\ and\ \bibinfo {editor}
  {\bibfnamefont {J.}~\bibnamefont {Millman}}}\ (\bibinfo {address} {Pasadena,
  CA},\ \bibinfo {year} {2008})\ pp.\ \bibinfo {pages} {11--15}\BibitemShut
  {NoStop}%
\bibitem [{\citenamefont {Br{\'e}laz}(1979)}]{Brelaz1979}%
  \BibitemOpen
  \bibfield  {author} {\bibinfo {author} {\bibfnamefont {D.}~\bibnamefont
  {Br{\'e}laz}},\ }\bibfield  {title} {\bibinfo {title} {New methods to color
  the vertices of a graph},\ }\href {https://doi.org/10.1145/359094.359101}
  {\bibfield  {journal} {\bibinfo  {journal} {Commun. ACM}\ }\textbf {\bibinfo
  {volume} {22}},\ \bibinfo {pages} {251} (\bibinfo {year} {1979})}\BibitemShut
  {NoStop}%
\bibitem [{\citenamefont {Powell}(1994)}]{Powell1994}%
  \BibitemOpen
  \bibfield  {author} {\bibinfo {author} {\bibfnamefont {M.~J.~D.}\
  \bibnamefont {Powell}},\ }\bibfield  {title} {\bibinfo {title} {A direct
  search optimization method that models the objective and constraint functions
  by linear interpolation},\ }in\ \href
  {https://doi.org/10.1007/978-94-015-8330-5_4} {\emph {\bibinfo {booktitle}
  {Advances in Optimization and Numerical Analysis}}},\ \bibinfo {editor}
  {edited by\ \bibinfo {editor} {\bibfnamefont {S.}~\bibnamefont {Gomez}}\ and\
  \bibinfo {editor} {\bibfnamefont {J.-P.}\ \bibnamefont {Hennart}}}\ (\bibinfo
   {publisher} {Kluwer Academic},\ \bibinfo {address} {Dordrecht},\ \bibinfo
  {year} {1994})\ pp.\ \bibinfo {pages} {51--67}\BibitemShut {NoStop}%
\bibitem [{\citenamefont {Javadi-Abhari}\ \emph {et~al.}(2024)\citenamefont
  {Javadi-Abhari}, \citenamefont {Treinish}, \citenamefont {Krsulich},
  \citenamefont {Wood}, \citenamefont {Lishman}, \citenamefont {Gacon},
  \citenamefont {Martiel}, \citenamefont {Nation}, \citenamefont {Bishop},
  \citenamefont {Cross}, \citenamefont {Johnson},\ and\ \citenamefont
  {Gambetta}}]{JavadiAbhari2024}%
  \BibitemOpen
  \bibfield  {author} {\bibinfo {author} {\bibfnamefont {A.}~\bibnamefont
  {Javadi-Abhari}}, \bibinfo {author} {\bibfnamefont {M.}~\bibnamefont
  {Treinish}}, \bibinfo {author} {\bibfnamefont {K.}~\bibnamefont {Krsulich}},
  \bibinfo {author} {\bibfnamefont {C.~J.}\ \bibnamefont {Wood}}, \bibinfo
  {author} {\bibfnamefont {J.}~\bibnamefont {Lishman}}, \bibinfo {author}
  {\bibfnamefont {J.}~\bibnamefont {Gacon}}, \bibinfo {author} {\bibfnamefont
  {S.}~\bibnamefont {Martiel}}, \bibinfo {author} {\bibfnamefont {P.~D.}\
  \bibnamefont {Nation}}, \bibinfo {author} {\bibfnamefont {L.~S.}\
  \bibnamefont {Bishop}}, \bibinfo {author} {\bibfnamefont {A.~W.}\
  \bibnamefont {Cross}}, \bibinfo {author} {\bibfnamefont {B.~R.}\ \bibnamefont
  {Johnson}},\ and\ \bibinfo {author} {\bibfnamefont {J.~M.}\ \bibnamefont
  {Gambetta}},\ }\bibfield  {title} {\bibinfo {title} {Quantum computing with
  {Qiskit}},\ }\href@noop {} {\bibfield  {journal} {\bibinfo  {journal}
  {arXiv:2405.08810}\ } (\bibinfo {year} {2024})}\BibitemShut {NoStop}%
\end{thebibliography}%

\end{document}